\documentclass[iop,revtex4]{emulateapj}
\usepackage{natbib}
\usepackage[normalem]{ulem}
\usepackage{color}
\definecolor{red}{rgb}{1.0,0.0,0.0}

\shorttitle{Spectroscopic characterization of HD 95086 \lowercase{b} with GPI}
\shortauthors{De Rosa et al.}

\tabletypesize{\normalsize}

\begin{document}

\title{Spectroscopic characterization of HD 95086 \lowercase{b} with the Gemini Planet Imager}

\author{Robert J. De Rosa\altaffilmark{1},
Julien Rameau\altaffilmark{2},
Jenny Patience\altaffilmark{3},
James R. Graham\altaffilmark{1},
Ren\'{e} Doyon\altaffilmark{2},
David Lafreni\`{e}re\altaffilmark{2},
Bruce Macintosh\altaffilmark{4},
Laurent Pueyo\altaffilmark{5},
Abhijith Rajan\altaffilmark{3},
Jason J. Wang\altaffilmark{1},
Kimberly Ward-Duong\altaffilmark{3},
Li-Wei Hung\altaffilmark{6},
J\'{e}r\^{o}me Maire\altaffilmark{7},
Eric L. Nielsen\altaffilmark{8,4},
S. Mark Ammons\altaffilmark{9},
Joanna Bulger\altaffilmark{10},
Andrew Cardwell\altaffilmark{11},
Jeffrey K. Chilcote\altaffilmark{7},
Ramon L. Galvez\altaffilmark{12},
Benjamin L. Gerard\altaffilmark{13},
Stephen Goodsell\altaffilmark{14,15},
Markus Hartung\altaffilmark{12},
Pascale Hibon\altaffilmark{16},
Patrick Ingraham\altaffilmark{17},
Mara Johnson-Groh\altaffilmark{13},
Paul Kalas\altaffilmark{1,8},
Quinn M. Konopacky\altaffilmark{18},
Franck Marchis\altaffilmark{8},
Christian Marois\altaffilmark{19,13},
Stanimir Metchev\altaffilmark{20,21},
Katie M. Morzinski\altaffilmark{22},
Rebecca Oppenheimer\altaffilmark{23},
Marshall D. Perrin\altaffilmark{5},
Fredrik T. Rantakyr\"{o}\altaffilmark{12},
Dmitry Savransky\altaffilmark{25},
Sandrine Thomas\altaffilmark{13}}

\altaffiltext{1}{Astronomy Department, University of California, Berkeley, CA 94720, USA}
\altaffiltext{2}{Institut de Recherche sur les Exoplan\`{e}tes, D\'{e}partment de Physique, Universit\'{e} de Montr\'{e}al, Montr\'{e}al QC H3C 3J7, Canada}
\altaffiltext{3}{School of Earth and Space Exploration, Arizona State University, PO Box 871404, Tempe, AZ 85287, USA}
\altaffiltext{4}{Kavli Institute for Particle Astrophysics and Cosmology, Stanford University, Stanford, CA 94305, USA}
\altaffiltext{5}{Space Telescope Science Institute, 3700 San Martin Drive, Baltimore, MD 21218, USA}
\altaffiltext{6}{Department of Physics and Astronomy, University of California Los Angeles, 430 Portola Plaza, Los Angeles, CA 90095, USA}
\altaffiltext{7}{Dunlap Institute for Astronomy and Astrophysics, University of Toronto, Toronto, ON, M5S 3H4, Canada}
\altaffiltext{8}{SETI Institute, Carl Sagan Center, 189 Bernardo Avenue, Mountain View, CA 94043, USA}
\altaffiltext{9}{Lawrence Livermore National Laboratory, L-210, 7000 East Avenue, Livermore, CA 94550, USA}
\altaffiltext{10}{Subaru Telescope, NAOJ, 650 North A'ohoku Place, Hilo, HI 96720, USA}
\altaffiltext{11}{Large Binocular Telescope Observatory, University of Arizona, 933 N. Cherry Ave, Tucson, AZ 85721, USA}
\altaffiltext{12}{Gemini Observatory, Casilla 603, La Serena, Chile}
\altaffiltext{13}{Department of Physics and Astronomy, University of Victoria, 3800 Finnerty Road, Victoria, BC, V8P 5C2, Canada}
\altaffiltext{14}{Department of Physics, Durham University, Stockton Road, Durham, DH1 3LE, UK}
\altaffiltext{15}{Gemini Observatory, 670 N. A'ohoku Place, Hilo, HI 96720, USA}
\altaffiltext{16}{European Southern Observatory, Alonso de Cordova 3107, Casilla 19001, Santiago, Chile}
\altaffiltext{17}{Large Synoptic Survey Telescope, 950 N. Cherry Ave, Tucson AZ 85719, USA}
\altaffiltext{18}{Center for Astrophysics and Space Science, University of California San Diego, La Jolla, CA 92093, USA}
\altaffiltext{19}{National Research Council of Canada Herzberg, 5071 West Saanich Road, Victoria, BC V9E 2E7, Canada}
\altaffiltext{20}{Department of Physics and Astronomy, Centre for Planetary Science and Exploration, The University of Western Ontario, London, ON N6A 3K7, Canada}
\altaffiltext{21}{Department of Physics and Astronomy, Stony Brook University, 100 Nicolls Road, Stony Brook, NY 11790, USA}
\altaffiltext{22}{Steward Observatory, 933 N. Cherry Avenue, University of Arizona, Tucson, AZ 85721, USA}
\altaffiltext{23}{American Museum of Natural History, New York, NY 10024, USA}
\altaffiltext{25}{Sibley School of Mechanical and Aerospace Engineering, Cornell University, Ithaca NY 14853}

\begin{abstract}
We present new $H$ (1.5--1.8\,\micron) photometric and $K_1$ (1.9--2.2\,\micron) spectroscopic observations of the young exoplanet HD~95086~b obtained with the Gemini Planet Imager. The $H$-band magnitude has been significantly improved relative to previous measurements, whereas the low resolution $K_1$ ($\lambda/\delta\lambda \approx 66$) spectrum is featureless within the measurement uncertainties, and presents a monotonically increasing pseudo-continuum consistent with a cloudy atmosphere. By combining these new measurements with literature $L^{\prime}$ photometry, we compare the spectral energy distribution of the planet to other young planetary-mass companions, field brown dwarfs, and to the predictions of grids of model atmospheres. HD~95086~b is over a magnitude redder in $K_1-L^{\prime}$ color than 2MASS~J12073346-3932539~b and HR~8799~c and d, despite having a similar $L^{\prime}$ magnitude. Considering only the near-infrared measurements, HD~95086~b is most analogous to the brown dwarfs 2MASS~J2244316+204343 and 2MASS~J21481633+4003594, both of which are thought to have dusty atmospheres. Morphologically, the spectral energy distribution of HD~95086~b is best fit by low temperature ($T_{\rm eff} =$ 800--1300~K), low surface gravity spectra from models which simulate high photospheric dust content. This range of effective temperatures is consistent with field L/T transition objects, but the spectral type of HD~95086~b is poorly constrained between early L and late T due to its unusual position the color-magnitude diagram, demonstrating the difficulty in spectral typing young, low surface gravity substellar objects. As one of the reddest such objects, HD~95086~b represents an important empirical benchmark against which our current understanding of the atmospheric properties of young extrasolar planets can be tested.
\end{abstract}


\section{Introduction}
Directly imaged exoplanets form an important subset of the exoplanet population for which it is possible to characterize atmospheric properties from the optical and thermal-infrared emission from the planet. Given the steep, monotonic decline in planet brightness with time (e.g. \citealp{Burrows:1997ua}), the currently known directly imaged exoplanets are companions to young stars ($<150$~Myr), enabling investigations into their early evolution \citep{Chauvin:2004cy,Lafreniere:2008jt,Marois:2008ei,Lagrange:2009hq,Delorme:2013bo,Bailey:2014et,Kraus:2014hk,Naud:2014jx,Gauza:2015fw,Macintosh:2015ew}. While these young exoplanets share many of the same atmospheric properties as isolated brown dwarfs recently identified in young moving groups (e.g., \citealp{Liu:2013gy,Gagne:2015kf}), differences in the elemental abundances of their atmospheres may provide evidence as to their formation mechanism (e.g. \citealp{Oberg:2011je,Konopacky:2013jv}). Atmospheres of young, directly-imaged planets also represent an important comparison to the transmission (e.g., \citealp{Sing:2011dn,Deming:2013ge}) and emission (e.g., \citealp{Knutson:2008gl,Bean:2013dg}) spectra of intensely irradiated planets transiting older stars. 

Multi-wavelength infrared photometry of imaged planet atmospheres (e.g., \citealp{Chauvin:2004cy,Marois:2008ei,Currie:2013cv,Skemer:2014hy}) has defined a sequence that is under-luminous and redder than field brown dwarfs of the same spectral type. Low resolution spectra of planetary mass companions (e.g., \citealp{Janson:2010db,Patience:2012cx,Barman:2011dq, Bonnefoy:2014bx}) exhibit signatures of low surface gravity and spectral features indicating the effects of clouds and non-equilibrium chemistry (e.g., \citealp{Hinz:2010fy,Barman:2011fe,Skemer:2012gr,Bonnefoy:2014bx}). Altogether, these properties are difficult to reproduce with state-of-the-art atmospheric models, which have difficulty explaining the discrepancies seen with respect to field L- and T-dwarfs.

The Scorpius-Centaurus association (Sco-Cen) is the nearest OB association \citep{deZeeuw:1999fe}, and has undergone multiple and complex phases of massive star formation, resulting in a large sample of young (5--20~Myr) and intermediate-mass (1.2--2.0~M$_\odot$) stars \citep{Mamajek:2002cl,Rizzuto:2011gs,Pecaut:2012gp,Song:2012gc}. Its well-constrained young age, proximity (90--150~pc), and the presence of near-infrared excess around a large number of its stars \citep{Chen:2012ki}---tracing leftover material from planet formation---make Sco-Cen ideal for giant planet searches with direct imaging. Thus far, five low-mass substellar companions ($M<25$~$M_{\rm Jup}$) have been detected around Sco-Cen stars: 1RXS~J160929.1-210524~b \citep{Lafreniere:2008jt}, GSC~06214-00210~b \citep{Ireland:2011id}, HIP~78530~B \citep{Lafreniere:2011dh}, HD~95086~b \citep{Rameau:2013dr}, and HD~106906~b \citep{Bailey:2014et}.

The star HD~95086 recently became a prime target to investigate the early phase of giant planet evolution. It is an A8, $1.7$~$M_\odot$ member of the Sco-Cen subgroup Lower Centaurus Crux \citep{deZeeuw:1999fe,Madsen:2002ha}, with an age of $17\pm4$~Myr \citep{Meshkat:2013fz} and at distance of $90.4\pm3.4$~pc \citep{vanLeeuwen:2007dc}. HD~95086 also harbors a massive debris disk, first traced by a large infrared excess \citep{Rizzuto:2012hx,Chen:2012ki} then resolved with \textit{Herschel}/PACS \citep{Moor:2013bg}. Modelling the spectral energy distribution (SED) and images of the system suggests a three component debris structure, like that of HR 8799 \citep{Su:2015ju}. With VLT/NaCo, \citet{Rameau:2013dr,Rameau:2013ds} discovered and confirmed the presence of a co-moving giant planet, HD~95086~b, at a projected separation of $56$~AU. This planet may be responsible for sculpting the outer edge of the debris gap \citep{Rameau:2013dr,Su:2015ju}. This discovery added another system to the relatively small population of young, nearby, intermediate-mass stars hosting both debris disks and directly imaged giant planets. 

The observed $L\,'$ ($3.8~\mu m$) luminosity of HD~95086~b indicated a ``hot-start'' model-dependent mass of $5\pm2$~$M_{\rm Jup}$ \citep{Rameau:2013dr}. ``Warm-start'' models predicted masses from 4--14~$M_{\rm Jup}$ for a wide range of initial entropy values (8--13~$k_{\rm B}/ \mathrm{baryon}$; \citealp{Galicher:2014er}). Follow-up observations with the Gemini Planet Imager at {\it H} ($1.65~\mu m$) and {\it K}$_1$ ($2.06~\mu m$) enabled initial constraints of its atmospheric properties \citep{Galicher:2014er}: the very red colors of $H-L\,' = 3.6 \pm 1.0$~mag and $K_1 - L\,' = 2.7 \pm 0.70$~mag suggested the atmosphere of HD~95086~b contains a high amount of photospheric dust, placing the planet at the L/T transition. Empirical comparisons of the position of HD~95086~b on the color-magnitude diagram showed that it is redder than dusty late field L-dwarfs \citep{Stephens:2009cc} and closer to, but still redder than, the young imaged planets HR~8799~cde \citep{Marois:2008ei,Marois:2010gp} and 2M1207~b \citep{Chauvin:2004cy}. Therefore, clouds may play an important role in the atmosphere of HD~95086~b. Finally, the predictions of the atmospheric properties of HD~95086~b from {\sc BT-Settl} \citep{Allard:2012fp} and LESIA \citep{Baudino:2015kh} atmospheric model fitting to its SED yielded $T_\mathrm{eff}=600$--1500~K and $\log g=2.1$--4.5 (cgs). Low surface gravity seemed to be favored which, combined with the enhanced dust content, might inhibit the formation of methane expected in this temperature range due to non-equilibrium $\mathrm{CO}/\mathrm{CH}_4$ chemistry \citep{Barman:2011fe,Zahnle:2014hl}.

In this paper, we present the first $K_1$ spectrum and revised $H$ photometry of HD~95086~b, obtained with the Gemini Planet Imager (GPI; \citealp{Macintosh:2014js}), which we synthesize with existing  $L^{\prime}$ photometry to constrain the SED of the planet (\S~\ref{sec:obs}). In \S~\ref{sec:empirical}, we compare the SED of HD~95086~b to that of other substellar companions in Sco-Cen, other benchmark planetary-mass companions, and field brown dwarfs. We also fit the full SED to several grids of model atmospheric spectra in \S~\ref{sec:models}, in an attempt to further constrain the atmospheric and physical properties of the planet. We also took advantage of this study to derive color transformations between the 2MASS, MKO, and GPI photometric systems as a function of spectral type.

\section{Observations and Data Reduction}\label{sec:obs}
\subsection{Observing sequence}
Integral field spectrograph observations of HD~95086 were made with GPI on 2015 April 8 UT at $K_1$ and 2016 February 29 UT at $H$ under program IDs GS-2015A-Q-501 and GS-2015B-Q-500. The $H$-band filter within GPI is similar to the MKO $H$ filter \citep{Tokunaga:2002ex}, while the $K_1$ filter spans the blue half of the $K$-band atmospheric window. The characterization of the $H$, $K_1$, and $K_2$ GPI filters is described in the Appendix. The observations of HD~95086 were timed such that the star transited the meridian during the observing sequence, maximizing the field rotation for Angular Differential Imaging (ADI; \citealp{Marois:2006df}) observations. Wavelength calibration measurements using an argon arc lamp were taken immediately following each sequence. At $K_1$, 42 $\times$ 119.3-second exposures were obtained over the course of two hours, achieving a field rotation of 39~degrees. Six sky measurements offset by $20\arcsec$ from the position of HD~95086 were also taken, three during the middle of the sequence and three at the end, to characterize the thermal emission from the sky, the telescope, and the instrument itself. Observing conditions were good, with an average DIMM seeing of $0\farcs 74$, MASS coherence time of $1.28$~ms, and wind speed of 0.93~m~s$^{-1}$. Of the 42 exposures, four were rejected due to poor image quality, and one due to the instrument shutter failing to open. At $H$, 37 $\times$ 59.6-second exposures were obtained over the course of an hour for a total rotation of 16.9~degrees. The observing conditions were excellent, with an average DIMM seeing of $0\farcs 94$, MASS coherence time of $4.7$~ms, and wind speed of 5.09~m~s$^{-1}$.

\subsection{Initial reduction steps}
The observations of HD~95086 were reduced using the standard primitives contained within the GPI Data Reduction Pipeline\footnote{\tt http://docs.planetimager.org/pipeline} (DRP; \citealp{Perrin:2014jh}) v1.3.0. The reduction process consisted of seven steps; (1) dark current subtraction; (2) bad pixel identification and interpolation; (3) compensation for instrument flexure between reference arcs taken several months previous, and the arcs taken during the observing sequence \citep{Wolff:2014cn}; (4) microspectra extraction to create an $(x, y, \lambda)$ data cube \citep{Maire:2014gs}; (5) division by a flat field to correct for lenslet throughput; (6) interpolation to a common wavelength scale; and (7) correction for geometric distortion of the image, as measured with a calibrated pinhole mask in the laboratory \citep{Konopacky:2014hf}. The interpolation step oversamples the final spectra, which have a spectral resolving power of $\lambda/\delta\lambda\approx66$ at $K_1$. This process was repeated for the sky measurements taken during the observing sequence at $K_1$, an average of which was subtracted from the reduced science data cubes. Images were then registered to a common center using the barycenter of the four satellite spots, which are attenuated replicas of the central point spread function (PSF) generated by a pupil plane diffraction grating, and for which the positions are measured by the GPI DRP \citep{Wang:2014ki}.

\subsection{Speckle noise minimization}
\begin{figure}
\epsscale{1.2}
\plotone{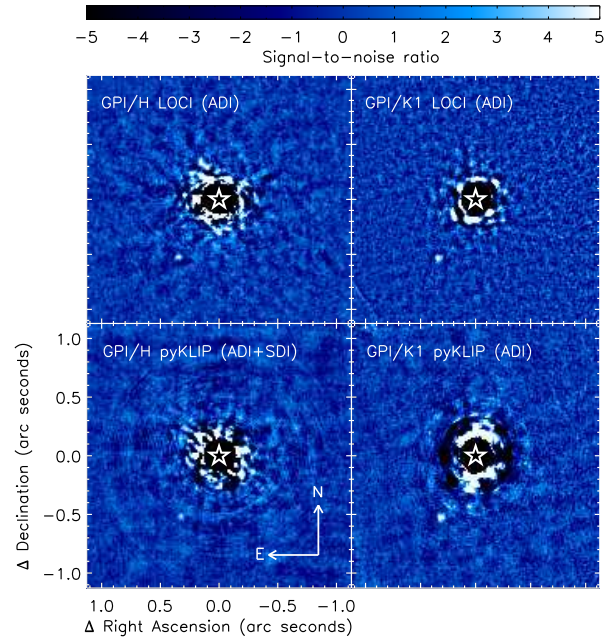}
\caption{SNR map of the HD~95086 observations taken at $H$ (left column) and $K_1$ (right column). Each dataset was processed with both the LOCI-based (top row) and pyKLIP (bottom row) pipelines. The SNR was calculated using a $1.5\lambda/D$ aperture. HD~95086~b was detected at a SNR of $\sim$5--7 at $H$ and $\sim 7$ at $K_1$. The central black mask conceals the region within which PSF subtraction is not performed, and the position of the central star position is designated by the star symbol.}
\label{fig:im}
\end{figure}
The reduced data cubes from each observing sequence were further processed through two independent pipelines using two speckle subtraction algorithms: the Karhunen-Lo\`{e}ve Image Projection algorithm (KLIP; \citealp{Soummer:2012ig}), and the Locally Optimized Combination of Images algorithm (LOCI; \citealp{Lafreniere:2007bg}). The speckle noise was minimized in each wavelength channel to create a residual spectral cube.

The KLIP-based pipeline processed the reduced data using pyKLIP\footnote{\tt https://bitbucket.org/pyKLIP/pyklip} \citep{Wang:2015th}, an open-source Python implementation of the KLIP algorithm. First, the individual wavelength slices from each reduced data cube were high-pass filtered in the Fourier domain to remove the low spatial frequency seeing halo. For the $K_1$ observations, the pipeline only utilized ADI due to the relatively small wavelength range of the $K_1$ filter, and in an attempt to avoid biases inherent to Spectral Differential Imaging (SDI; \citealp{Marois:2000jt}). PSF subtraction was performed over six annuli of width 13~pixels ($0\farcs 184$), starting from an inner separation of 11~pixels ($0\farcs 156$). Each annulus was subdivided into four 90~degree arcs. A PSF constructed from the projection of the individual images onto the first five Karhunen-Lo\`{e}ve modes was subtracted from each region. The PSF-subtracted images were de-rotated and averaged together. HD~95086~b was detected at a SNR of $\sim$5 and $\sim$7 in the collapsed $H$ and $K_1$ broadband images, which are shown in Figure \ref{fig:im}.

For the LOCI-based pipeline, the low spatial frequency signal was removed from each image by subtracting a median-filtered version of the image, using a 15 pixel wide box. The speckle subtraction was then performed using LOCI with subtraction annuli of width five pixels, optimization regions of $N_a=300$ full width at half maximum (FWHM, $\simeq 3.5/3.9$~px at $H/K_1$) with a geometry factor of $g=1$, and a separation criteria of $N_\delta=0.75$ FWHM. In each case, the parameters are as defined in \citet{Lafreniere:2007bg}. Each residual image was then derotated, mean combined with a trimmed mean ($10\%$), and aligned north up to generate each final residual slice. The final residual images within each filter are shown in Figure \ref{fig:im}. At $K_1$, HD~95086~b is detected at an SNR of 3--5 in individual channels, and at an SNR of $\sim$7.5 in the broadband image. At $H$, HD~95086~b was recovered in the broadband image with a similar SNR. The astrometry of HD~95086~b will be discussed in a seperate study (Rameau et al. 2016, {\it submitted})

\subsection{Spectrophotometry extraction}
\begin{deluxetable}{cccc}
\tablecaption{Contrast between and spectral energy distribution of HD~95086 and HD~95086~b, with corresponding 68\% confidence intervals.}
\tablewidth{0pt}
\tablecolumns{4}
\tablehead{
\colhead{$\lambda_{\rm eff}$} & \colhead{Contrast} & \colhead{$F_{\lambda}({
\rm A}) \times 10^{-13}$} & \colhead{$F_{\lambda}({\rm b}) \times 10^{-18}$} \\
\colhead{($\micron$)}  &  \colhead{($10^{-6}$)} & \multicolumn{2}{c}{(W~m$^{-2}$~$\micron^{-1}$)}}
\startdata
1.6330 & $3.31_{-0.68}^{+0.86}$ & $21.79\pm0.23$ & $7.22_{-1.49}^{+1.87}$\\
\vdots & \vdots & \vdots\\
 1.9548 & \phantom{0}$4.78\pm4.30$ & $11.54\pm0.13$ & \phantom{0}$5.51\pm4.96$\\
1.9634 & \phantom{0}$4.65\pm3.53$ & $11.57\pm0.13$ & \phantom{0}$5.38\pm4.08$\\
1.9719 & \phantom{0}$6.99\pm3.22$ & $11.54\pm0.12$ & \phantom{0}$8.07\pm3.72$\\
1.9805 & \phantom{0}$5.66\pm2.77$ & $11.44\pm0.12$ & \phantom{0}$6.47\pm3.17$\\
1.9891 & \phantom{0}$8.32\pm2.61$ & $11.29\pm0.12$ & \phantom{0}$9.38\pm2.95$\\
1.9977 & \phantom{0}$8.63\pm2.44$ & $11.12\pm0.12$ & \phantom{0}$9.59\pm2.71$\\
2.0063 & \phantom{0}$8.45\pm2.60$ & $10.95\pm0.12$ & \phantom{0}$9.25\pm2.85$\\
2.0149 & \phantom{0}$9.99\pm3.10$ & $10.77\pm0.11$ & $10.77\pm3.34$\\
2.0234 & \phantom{0}$9.84\pm3.35$ & $10.60\pm0.11$ & $10.43\pm3.55$\\
2.0320 & $10.78\pm2.66$ & $10.43\pm0.11$ & $11.24\pm2.78$\\
2.0406 & $10.72\pm2.40$ & $10.27\pm0.11$ & $11.01\pm2.46$\\
2.0492 & \phantom{0}$9.51\pm2.30$ & $10.11\pm0.11$ & \phantom{0}$9.61\pm2.33$\\
2.0578 & \phantom{0}$9.98\pm2.59$ & \phantom{0}$9.95\pm0.11$ & \phantom{0}$9.93\pm2.58$\\
2.0664 & $10.73\pm2.70$ & \phantom{0}$9.79\pm0.10$ & $10.51\pm2.65$\\
2.0749 & \phantom{0}$9.55\pm2.46$ & \phantom{0}$9.64\pm0.10$ & \phantom{0}$9.21\pm2.37$\\
2.0835 & \phantom{0}$9.92\pm2.43$ & \phantom{0}$9.50\pm0.10$ & \phantom{0}$9.42\pm2.31$\\
2.0921 & $10.83\pm2.63$ & \phantom{0}$9.35\pm0.10$ & $10.13\pm2.46$\\
2.1007 & $10.34\pm2.71$ & \phantom{0}$9.20\pm0.10$ & \phantom{0}$9.51\pm2.49$\\
2.1093 & $13.58\pm2.73$ & \phantom{0}$9.06\pm0.10$ & $12.30\pm2.47$\\
2.1179 & $12.81\pm2.84$ & \phantom{0}$8.91\pm0.10$ & $11.41\pm2.53$\\
2.1264 & $10.88\pm2.79$ & \phantom{0}$8.77\pm0.09$ & \phantom{0}$9.53\pm2.44$\\
2.1350 & $13.88\pm2.79$ & \phantom{0}$8.61\pm0.09$ & $11.95\pm2.40$\\
2.1436 & $14.69\pm3.45$ & \phantom{0}$8.41\pm0.09$ & $12.36\pm2.90$\\
2.1522 & $14.53\pm3.69$ & \phantom{0}$8.17\pm0.09$ & $11.86\pm3.01$\\
2.1608 & $18.99\pm4.19$ & \phantom{0}$7.92\pm0.09$ & $15.04\pm3.32$\\
2.1694 & $18.36\pm5.17$ & \phantom{0}$7.78\pm0.09$ & $14.29\pm4.02$\\
2.1779 & $16.76\pm5.82$ & \phantom{0}$7.77\pm0.09$ & $13.02\pm4.52$\\
2.1865 & $17.42\pm6.58$ & \phantom{0}$7.78\pm0.08$ & $13.56\pm5.13$\\
2.1951 & $17.02\pm9.91$ & \phantom{0}$7.76\pm0.08$ & $13.20\pm7.68$\\
\vdots & \vdots & \vdots\\
3.7697 & $161.42_{-25.91}^{+30.88}$ & $0.99\pm0.01$ & $15.95_{-2.56}^{+3.05}$
\enddata
\label{tab:sed}
\end{deluxetable}
\begin{figure}
\epsscale{1.2}
\plotone{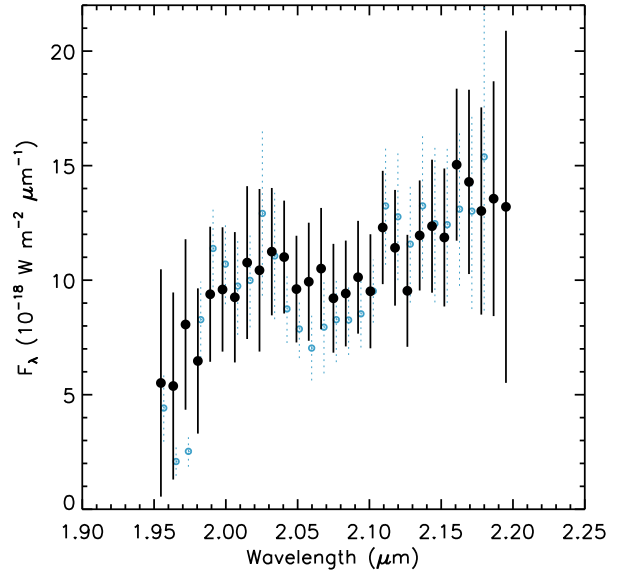}
\caption{GPI $K_1$ spectrum of HD~95086~b extracted from the LOCI-based (filled black circles) and pyKLIP (open blue circles) pipelines. For clarity the pyKLIP spectrum has been slightly offset in the wavelength direction. The first eight channels (1.886--1.946\micron) were excluded due to the poor quality of the extracted spectrum at those wavelengths. The covariance between neighbouring spectral channels in the adopted spectrum is described in Section~\ref{sec:covariance}.}
\label{fig:spectrum}
\end{figure}

The $K_1$ spectrum of HD~95086~b was extracted from the final PSF-subtracted image of each pipeline using two different techniques. For the KLIP-based pipeline, the spectrum was extracted from each wavelength slice using a $1.5\lambda/D$ aperture, with a center based on the position of the planet in the wavelength-collapsed image. Uncertainties were estimated by measuring the noise within an annulus of width $1.5\lambda/D$, and radius equivalent to the star to planet separation. In order to account for self-subtraction within the KLIP algorithm, the throughput was calculated by injecting a copy of the average of the four satellite spots, scaled to a similar brightness to HD~90586~b, into the data cubes prior to PSF subtraction. This copy was injected at the same separation as HD~95086~b. The throughput was then estimated at each wavelength by comparing the flux of the injected source to the flux measured in the PSF-subtracted image. This process was repeated 25 times, with the position angle of the injected source rotated for each trial. For each wavelength, the average throughput was calculated from these trials, and the standard deviation was used as the uncertainty on the throughput. To correct for throughput, the flux of the planet at a given wavelength was scaled by the average throughput, and the uncertainties on the flux and throughput were combined in quadrature. The flux from the central star was estimated by performing aperture photometry on an average of the four satellite spots within each wavelength slice prior to PSF subtraction, multiplied by the satellite spot to star flux ratio measured in the laboratory to be $\Delta m=8.92\pm0.06$~mag \citep{Maire:2014gs}. The contrast between star and planet was then calculated as the ratio of their fluxes.

The LOCI-based pipeline first derived the position and contrast of HD~95086~b from the broadband image---created by averaging the 37 wavelength channels of the final PSF-subtracted data cube---using the fake planet injection technique (e.g., \citealp{Lagrange:2010fs}). A PSF was generated from the average of the four satellite spots prior to PSF subtraction, integrated over the band and averaged over the temporal sequence, which was scaled and subtracted from the data at the estimated flux and position of the planet before performing the LOCI-based PSF subtraction. The flux and the position of the planet were iterated using an amoeba-simplex optimization algorithm \citep{Nelder:1965in} until the sum of the squared pixel intensities in an arc of size $2\times4$ FWHM was minimized. Uncertainties on the position and contrast were estimated from the root mean square of fake planets injected with the measured contrasts and separations of HD~95086~b at ten different position angles uniformly distributed between 90 and 270 deg away from the planet. These injected planets were then characterized using the same strategy as for HD~95086~b \citep[e.g.,][]{Marois:2010gp, Lagrange:2010fs, Macintosh:2015ew}. From the $K_1$ observations, a revised contrast of $\Delta K_1=12.20\pm0.20$ was measured for HD~95086~b, the uncertainty on which was calculated by combining in quadrature the star-to-spot ratio uncertainty ($0.07$ mag), the PSF variability ($0.07$ mag), and the measurement error ($0.18$ mag). The $H$-band contrast was similarly measured as $\Delta H=13.70\pm0.25$, the uncertainty on which combines the star-to-spot ratio uncertainty ($0.06$ mag), PSF variability ($0.05$ mag), and measurement error ($0.24$ mag).

Because the SNR of HD~95086~b is lower within the individual slices of the PSF-subtracted $K_1$ data cube, the contrast at each wavelength---and their associated errors---were extracted using the same technique as described above, except that the position of the planet was fixed to that measured within the broadband image. Because LOCI biases due to over-subtraction can be difficult to calibrate, we also extracted the spectrum using the cADI algorithm \citep{Marois:2006df}. While the contrasts measured using LOCI and cADI were very similar, HD~95086~b was recovered in fewer wavelength channels in the cADI reduction. Although the $K_1$ spectrum of HD~95086~b extracted from the pyKLIP and LOCI-based pipelines agreed within the uncertainties (Fig. \ref{fig:spectrum}), the spectrum extracted from the LOCI-based pipeline was used for the remainder of this study due to the smaller point-to-point scatter. The adopted contrasts measured within the $K_1$ bandpass between the star and planet are given in Table~\ref{tab:sed}. 

\label{sec:primaryfit}
The $K_1$ spectrum of HD~95086~b was then calculated by multiplying the contrasts by the predicted flux of HD~95086 at each wavelength. As no flux-calibrated $K_1$ spectrum of HD~95086 exists, a model stellar atmosphere from the {\sc BT-NextGen} grid\footnote{{\tt https://phoenix.ens-lyon.fr/Grids/BT-NextGen/SPECTRA/}} was used \citep{Allard:2012fp}. The effective temperature, surface gravity, and metallicity of HD~95086 were fit using optical broadband \citep{Hog:2000wk} and intermediate-band \citep{Rufener:1988wq} photometry, and broadband near-infrared \citep{Skrutskie:2006hla} and thermal-infrared \citep{Cutri:2014wx} photometry. By performing linear interpolation of the predicted flux for each filter between the model grid points, the best fit atmosphere was found with $T_{\rm eff} = 7633\pm24$~K, $\log g = 3.93\pm0.05$, $[M/H]=-0.35\pm0.01$, and a dilution factor of $R^2/d^2 = 10^{-18.8\pm0.01}$, where $R$ and $d$ are the radius of and distance to HD~95086. These values are consistent with previous literature estimates (e.g. \citealp{AllendePrieto:1999td, Wright:2003gs, Bonneau:2006bl, McDonald:2012cg}). The apparent $K_1$ spectrum of HD~95086 was estimated from the best fitting interpolated model, and is given in Table~\ref{tab:sed}. The GPI $H$ and $K_1$, and NaCo $L^{\prime}$ apparent magnitudes of HD~95086 were also estimated from the same model atmosphere using the filter transmission profiles discussed in the Appendix, and are given in Table~\ref{tab:properties}. The final $K_1$ spectrum of HD~95086~b is plotted in Figure~\ref{fig:spectrum}, and is given in Table~\ref{tab:sed}, in addition to the apparent fluxes calculated from the revised $H$-band contrast and the $L^{\prime}$ contrast from \citet{Galicher:2014er}.

\subsection{Spectral covariances}
\label{sec:covariance}
To properly assess the uncertainties on any fit to the $K_1$ spectrum of HD~95086~b, the covariances between neighbouring wavelength channels caused by both residual speckle noise in the final PSF-subtracted image, and the oversampling of the microspectra in the extraction process, must be taken into account. Following the method described in \citet{Greco:2016ww}, the correlation $\psi_{ij}$ between pixel values at wavelengths $\lambda_i$ and $\lambda_j$ within an annulus of width 1.5~$\lambda/D$ at a given separation was estimated as
\begin{equation}
    \psi_{ij} = \frac{\langle I_i I_j\rangle}{\sqrt{\langle I^2_i \rangle \langle I^2_j \rangle}}
\end{equation}
where $\langle I_i\rangle$ is the average intensity within the annulus at wavelength $\lambda_i$. This was repeated for all pairs of wavelengths, and at separations of $\rho=$ 400, 500, 620 (the separation of HD~95086~b), 700, 800, 900, and 1000~mas. The companion was masked in the 620~mas annulus. \citet{Greco:2016ww} parametrize the correlation $\psi_{ij}$ into three terms as
\begin{eqnarray}
\label{eqn:model}
    \psi_{ij}\approx A_{\rho}\exp\left[-\frac{1}{2}\left(\frac{\rho}{\sigma_{\rho}}\frac{\lambda_i - \lambda_j}{\lambda_c}\right)^2\right] \nonumber\\ + A_{\lambda}\exp\left[-\frac{1}{2}\left(\frac{1}{\sigma_{\lambda}}\frac{\lambda_i - \lambda_j}{\lambda_c}\right)^2\right] + A_{\delta}\delta_{ij}
\end{eqnarray}
where the symbols are as in \citet{Greco:2016ww}. The first two terms are correlated; the first models the contribution of speckle noise, while the second models other spectral correlations, such as correlations induced by the interpolation of the micro-spectra in the reduction process. The third term models uncorrelated noise, and does not contribute to the off-diagonal elements of the correlation matrix.

The measurements of $\psi_{ij}$ at the seven angular separations within the final PSF-subtracted image of HD~95086 were used to fit the parameters of the model in Equation~\ref{eqn:model}. The amplitudes of the three noise terms ($A_{\rho}$, $A_{\lambda}$, $A_{\delta}$) were allowed to vary as a function of angular separation, while the two correlation lengths ($\sigma_{\rho}$, $\sigma_{\lambda}$) were fixed over the entire field-of-view. As the sum of the three amplitudes was fixed at unity, there were a total of sixteen free parameters to fit. Due to the high dimensionality of the problem, a parallel-tempered Markov Chain Monte Carlo algorithm was used to properly sample the posterior distributions of each free parameter and to find the global minimum \citep{ForemanMackey:2013io}. The best fit parameters at the separation of HD~95086~b were found to be: $A_{\rho} = 0.077$, $A_{\lambda} = 0.851$, $A_{\delta} = 0.073$, $\sigma_{\rho} = 0.375$ $\lambda/D$, $\sigma_{\lambda} = 0.004$ $\lambda/D$. From this model of the correlation, the off-diagonal elements of the covariance matrix $C$ were then calculated using
\begin{equation}
    \psi_{ij} \equiv \frac{C_{ij}}{\sqrt{C_{ii}C_{jj}}}
\end{equation}
where the diagonal elements of $C$ were the square of the stated uncertainties on the spectrum of the planet given in Table~\ref{tab:sed}. Based on this analysis, the spectrum was shown to be highly correlated for flux measurements separated by up to two wavelength channels (each channel is $0.0086$~\micron~wide), with a significantly lower correlation for fluxes separated by more than two wavelength channels. Both the parametrized model and a visual inspection of the final images showed that the data were not dominated by speckles, such that the correlation can be ascribed almost entirely to the wavelength interpolation carried out as a part of the standard reduction procedure for GPI data.

\section{Results and Empirical Comparisons}
\label{sec:empirical}
\begin{deluxetable}{lccc}
\tablecaption{Properties of the HD 95086 system}
\tablewidth{0pt}
\tablehead{
\colhead{Property} & \multicolumn{2}{c}{Value} & \colhead{Unit}}
\startdata
Parallax                & \multicolumn{2}{c}{$11.06\pm0.41$\tablenotemark{\it a}} & mas\\
Distance                & \multicolumn{2}{c}{$90.42\pm3.35$\tablenotemark{\it a}} & pc\\
$\mu_{\alpha}$          & \multicolumn{2}{c}{$-41.41\pm0.42$\tablenotemark{\it a}} & mas yr$^{-1}$\\
$\mu_{\delta}$          & \multicolumn{2}{c}{$12.47\pm0.36$\tablenotemark{\it a}} & mas yr$^{-1}$\\
Age                     & \multicolumn{2}{c}{$17\pm4$\tablenotemark{\it b}} & Myr \\
\tableline
&HD 95086 & HD 95086 b\\
\tableline 
GPI $\Delta H$            & -- & $13.70 \pm 0.25$ & mag \\
GPI $\Delta K_1$          & -- & $12.20 \pm 0.20$ & mag \\
NaCo $\Delta L^{\prime}$  & -- & $9.48 \pm 0.19$\tablenotemark{\it c} & mag \\
2MASS $H$               & $6.867\pm0.047$\tablenotemark{\it d}            & -- & mag \\
GPI $H$                 & $6.807\pm0.025$\tablenotemark{\it e} & $20.51\pm0.25$ & mag \\
2MASS $K_{\rm S}$       & $6.789\pm0.021$\tablenotemark{\it d} & -- & mag \\
GPI $K_1$               & $6.785\pm0.025$\tablenotemark{\it e} & $18.99\pm0.20$ & mag \\
{\it WISE} $W1$         & $6.717\pm0.067$\tablenotemark{\it f} & -- & mag \\
NaCo $L\,'$             & $6.787\pm0.025$\tablenotemark{\it e} & $16.27\pm0.19$ & mag \\
 $H-K_1$             &  $0.022 \pm 0.035$   & $1.52 \pm 0.32$  & mag \\
$H-L\,'$            &  $0.019 \pm 0.035$   & $4.24 \pm 0.32$  & mag  \\
$K_1-L\,'$          &  $0.035 \pm 0.035$   & $2.72 \pm 0.28$  & mag \\
$M_H$                     & $2.025 \pm 0.084$   & $15.73 \pm 0.26$  & mag \\
$M_{K_1}$                 & $2.004 \pm 0.084$   & $14.20 \pm 0.22$ & mag \\
$M_{L\,'}$                & $2.006 \pm 0.084$   & $11.49 \pm 0.21$ & mag \\
SpT                 &  A$8\pm1$\tablenotemark{\it b}   & L1--T3  &  --   \\
$T_{\rm eff}$           & $7609 \pm 128$\tablenotemark{\it g} & 800--1300 & K \\
                        & $7633 \pm 24$\tablenotemark{\it e} & & K \\
 $\log g$               & $\sim4.15$\tablenotemark{\it h} & $\lesssim4.5$  & dex \\
                        & $3.93\pm0.05$\tablenotemark{\it e} & & dex\\
 $\log L/L_\odot$      &   $0.836\pm0.035$\tablenotemark{\it b}   &  $-4.96\pm0.10$\tablenotemark{\it i}   & dex \\
Mass                    & $\sim1.7$\tablenotemark{\it b}  & -- & $M_\odot$ \\
                       & -- & $4.4\pm0.8$\tablenotemark{\it j} & $M_{\rm Jup}$
\enddata
\tablecomments{a -- \citet{vanLeeuwen:2007dc}, b -- \citet{Meshkat:2013fz}, c -- \citet{Galicher:2014er}, d -- \citet{Skrutskie:2006hla}, e -- Estimated from best fit {\sc BT-NextGen} model atmosphere (\S \ref{sec:primaryfit}), f --\citet{Cutri:2014wx}, g -- Based on an average of the measurements from \citet{AllendePrieto:1999td, Wright:2003gs, Bonneau:2006bl, McDonald:2012cg}, h -- Using Dartmouth isochrones from \citet{Dotter:2008ga}, i -- Using $K$-band bolometric correction, j -- Based on \citet{Baraffe:2003bj} evolutionary models}
\label{tab:properties}
\end{deluxetable}

\subsection{Photometry}
\label{sec:ccd}
We measured contrasts of $\Delta H=13.70\pm0.25$ and $\Delta K_1=12.20\pm0.20~$mag for HD\,95086\,b. These new measurements are in agreement with the contrasts given in \citet{Galicher:2014er}, but the uncertainties have been significantly reduced due to an improvement of the SNR of the detections. These contrasts, and the $L^{\prime}$ contrast given in \citet{Galicher:2014er}, were combined with the predicted flux of HD~95086 in each filter reported in Section~\ref{sec:primaryfit} to calculate the apparent magnitudes of HD~95086~b. Absolute magnitudes were then calculated using these apparent magnitudes, and the {\it Hipparcos} parallax of HD~95086. A summary of all available photometry of HD~95086~b is given in Table~\ref{tab:properties}.

\begin{figure}
\epsscale{1.2}
\plotone{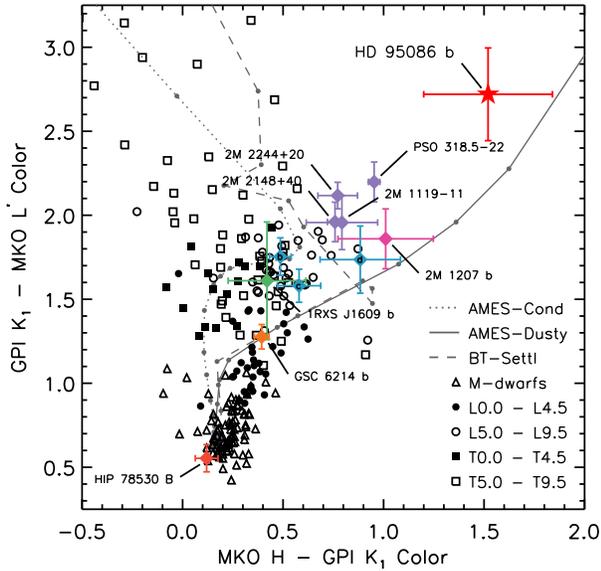}
\caption{MKO $H$ $-$ GPI $K_1$ vs. GPI $K_1$ $-$ MKO $L^{\prime}$ color-color diagram showing the position of HD~95086~b (red filled star) with respect to Sco-Cen substellar companions, benchmark planetary-mass companions, isolated young and/or dusty brown dwarfs, and field M-, L-, and T-dwarfs. HR~8799~bcd are plotted as blue filled circles, and are distinguished by a small letter within the symbol. Sources for the photometric measurements of the comparison objects are given in \S \ref{sec:ccd}. The colors of HD~95086~b were calculated using the GPI $H$-band and NaCo $L^{\prime}$ photometry, as discussed in \S \ref{sec:ccd}. For comparison, the 17~Myr isochrone was computed using the {\sc AMES-Cond} (dotted line), {\sc AMES-Dusty} (solid line), and {\sc BT-Settl} (dashed line) atmospheric models, in conjunction with the  \citet{Baraffe:2003bj} evolutionary models.}
\label{fig:CCD}
\end{figure}
The location of HD~95086~b on a $K_1-L^{\prime}$ vs. $H-K_1$  color-color diagram (CCD) is shown in Figure~\ref{fig:CCD}, alongside Sco-Cen substellar companions, benchmark planetary-mass companions, and isolated young and/or dusty brown dwarfs (listed in Table~\ref{tab:obj_comp}), and field M-dwarfs and brown-dwarfs within both the SpeX Prism Spectral Library\footnote{\tt http://pono.ucsd.edu/\char`\~adam/browndwarfs/spexprism} and those listed in \citet{Dupuy:2012bp}. As no spectral information exists within either the $H$- or $L$-band for HD~95086~b, a color transformation could not be determined between either GPI~$H$ and MKO~$H$ or NaCo~$L^{\prime}$ and MKO~$L^{\prime}$. Based on the empirical color transformations given in Table \ref{tab:filters}, the $H$-band color transformation is estimated to be small ($<0.1$~mag), despite the spectral type of HD~95086~b being relatively unconstrained (see \S \ref{sec:sed}). No empirical estimate for the $L$-band color transformation was made given the paucity of thermal-infrared spectra of brown dwarfs, although it is likely to be small relative to the size of the measurement uncertainties.

Photometry for the comparison objects was collated from a variety of sources: \citet{Lafreniere:2008jt} and \citet{Lachapelle:2015cx} for 1RXS~J160929.1-210524~b (1RXS~J1609~b); \citet{Bailey:2014et} for HD~106906~b; \citet{Ireland:2011id} and \citet{Lachapelle:2015cx} for GSC~06214-00210~b (GSC~6214~b); \citet{Bailey:2013gl} and \citet{Lachapelle:2015cx} for HIP~78530~B \citep{Lafreniere:2011dh}; \citet{Chauvin:2004cy} for 2MASS~J120734-393253~b (2M~1207~b); \citet{Marois:2008ei} and \citet{Esposito:2013hs} for HR~8799~bcd; \citet{Skrutskie:2006hla} and \citet{Cutri:2014wx} for 2MASS~J11193254-1137466 (2M~1119-11; \citealp{Kellogg:2015cf}), 2MASS~J21481628+4003593 (2M~2148+40; \citealp{Looper:2008hs}), and 2MASS~J22443167+2043433 (2M~2244+20; \citealp{Dahn:2002fu}); and \citet{Liu:2013gy} for PSO~J318.5338-22.8603 (PSO~318.5-22).

The MKO~$K$-band ($K$, $K_{\rm S}$, $K^{\prime}$) or 2MASS~$K_{\rm S}$ magnitudes for each comparison object were converted into GPI~$K_1$ magnitudes using their $K$-band spectra (see \S~\ref{sec:comparison_to_young}) using the procedure described in the Appendix. For the M-dwarfs and brown dwarfs within the SpeX library, the spectra were flux calibrated using either MKO or 2MASS $H$-band absolute magnitudes from \citet{Dupuy:2012bp}. The GPI $K_1$ magnitude was then estimated using this flux calibrated spectrum, and the $K_1$ filter profile, as described in the Appendix. For brown dwarfs within \citet{Dupuy:2012bp} without a spectrum within the SpeX library, the empirical color transformation given in Table~\ref{tab:filters} was used. MKO $L^{\prime}$ photometry for these objects was obtained from a variety of sources (\citealp{Dupuy:2012bp}, and references therein). If no MKO $L^{\prime}$ measurement was available, it was instead estimated from the {\it WISE}~$W1$ magnitude, and an empirical MKO~$L^{\prime}$ $-$ {\it WISE}~$W1$ color transformation given in the Appendix (Table~\ref{tab:filters}).

HD~95086~b was already known to have unusually red $H-L^{\prime}$ colors \citep{Meshkat:2013fz, Galicher:2014er}. Figure~\ref{fig:CCD} further demonstrates the peculiar nature of this object. It has a $K_1-L^{\prime}$ consistent with some late T-dwarfs, but an $H-K_1$ that is redder than known field objects. Relative to other young substellar objects, HD~95086~b has a $K_1-L^{\prime}$ and $H-K_1$ color that is approximately a magnitude redder than that of 2M~1207~b and HR~8799~d. Of the four isolated brown dwarfs identified as being young or possessing unusually dusty atmospheres, HD~95086~b is most analogous to PSO 318.5-22, one of the reddest field brown dwarfs known \citep{Liu:2013gy}, although HD~95086~b is half a magnitude redder in both $K_1-L^{\prime}$ and $H-K_1$. When compared with the location of the 17 Myr isochrones calculated from three different types of the atmospheric models, HD~95086~b is closer to the predicted colors of the models which simulate a large photospheric dust content ({\sc AMES-Dusty}, \citealp{Chabrier:2000hq,Allard:2001fh}). The location of these isochrones on the CCD does not vary significantly within the uncertainty of the age of HD~95086.

\begin{figure}
\epsscale{1.2}
\plotone{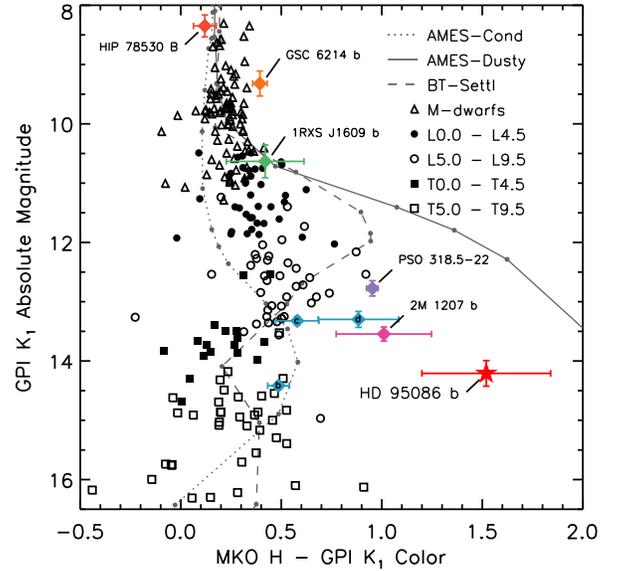}
\caption{GPI $M_{K_1}$ vs. MKO $H$ $-$ GPI $K_1$ color-magnitude diagram showing the position of HD~95086~b relative to substellar companions within Sco-Cen, benchmark planetary-mass companions, and field M-, L-, and T-dwarfs. Symbols and lines have the same meaning as in Figure~\ref{fig:CCD}.}
\label{fig:CMDa}
\end{figure}
\begin{figure}
\epsscale{1.2}
\plotone{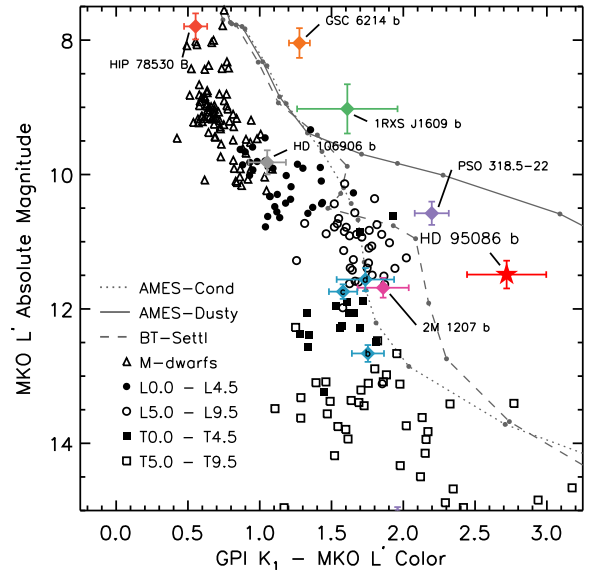}
\caption{MKO $L^{\prime}$ vs. GPI $K_1$ $-$ MKO $L^{\prime}$ color-magnitude diagram showing the position of HD~95086~b relative to substellar companions within Sco-Cen, benchmark planetary-mass companions, and field M-, L-, and T-dwarfs. Symbols and lines have the same meaning as in Figure~\ref{fig:CCD}. The extremely red $K_1-L^{\prime}$ color is more consistent with field late T-dwarfs, while the $L^{\prime}$ absolute magnitude suggests an earlier spectral type.}
\label{fig:CMDb}
\end{figure}

The updated photometry of HD~95086~b was also used to place it on two color-magnitude diagrams (CMDs); $M_{K_1}$ vs. $H-K_1$ in Figure~\ref{fig:CMDa}, and $M_{L^{\prime}}$ vs. $K_1-L^{\prime}$ in Figure~\ref{fig:CMDb}. In both diagrams, the Sco-Cen substellar companions are both brighter and bluer than HD~95086~b, consistent with the planet being cooler, less massive, and with a dustier photosphere. HD~95086~b has a $K_1$ luminosity consistent with mid-T dwarfs, and an $L^{\prime}$ luminosity consistent with L/T transition objects. In both instances, the planet is significantly redder than the field population, and lies on an apparent extension to the L-dwarf sequence. This extension is populated with lower surface gravity objects that have maintained cloudy atmospheres at temperatures for which higher-surface gravity field brown dwarfs have transitioned to clear atmospheres \citep{Kirkpatrick:2005cv}. As previously recognized \citep{Galicher:2014er}, HD~95086~b has a similar absolute $L^{\prime}$ magnitude to HR~8799~cd and 2M~1207~b, suggesting similar effective temperatures, but it is approximately a magnitude fainer in $K_1$, with a similar luminosity to HR~8799~b.

Finally, the very red $K_1-L^{\prime}$ color of HD~95086~b could also be explained by the presence of circumplanetary material, either a dust shell or a disc, as is the case for GSC~6214~b \citep{Bailey:2013gl}. Contrary to GSC~6214~b, which has strong evidence of accretion \citep{Bowler:2014,Zhou:2014ct,Lachapelle:2015cx}, no additional constraints beyond the red color support this hypothesis for HD~95086~b. Improvements in atmospheric modelling and the treatment of dust were able to explain the underluminosity of 2M~1207~b \citep{Barman:2011dq}, which had previously been suggested was due to the presence of circumstellar material \citep{Mohanty:2007er}. Future narrow-band observations sensitive to, for example, H$\alpha$ or Br$\gamma$ emission are required in order to further test this hypothesis.

\subsection{Spectroscopy and the full SED}\label{sec:sed}
\label{sec:spectrum}
The $K_1$ spectrum of HD~95086~b shown in Figure~\ref{fig:spectrum} exhibits a relatively featureless slope rising towards longer wavelengths, as is seen for other young and/or dusty brown dwarfs (e.g., \citealp{Stephens:2009cc, Patience:2010hf, Allers:2013hk, Liu:2013gy}). There is no evidence of strong molecular absorption due to CH$_4$ and H$_2$O beyond 2.08~\micron~\citep{Kirkpatrick:2005cv}, as is seen within the cloud-free atmospheres of mid to late-T dwarfs (e.g., \citealp{Naud:2014jx,Gagne:2015kf}). The spectrum of HD~95086~b does show a slight decrease in flux between $2.03$ and $2.12$~\micron. However, given the size of the uncertainties on the spectrum relative to the size of the decrease, it is unclear whether this is an absorption feature within the spectrum, or the result of the low SNR of the spectrum.

\subsubsection{Spectral type \& comparison to field objects}
\label{sec:spex_comparison}
To assess the spectral type of HD~90586~b, the literature photometry and the new $K_1$ spectrum were compared with stars and brown dwarfs within the SpeX Prism Spectral Library and the IRTF Spectral Library \citep{Cushing:2005ed,Rayner:2009ki}. As most objects within both spectral libraries had no $L$-band spectroscopy from which a synthetic magnitude could be calculated, MKO $L^{\prime}$ photometric measurements were obtained from a number of literature sources \citep{Stephens:2001iz, Leggett:2003gt, Golimowski:2004en, Leggett:2010cl, Dupuy:2012bp, Faherty:2012cy}. For those objects without literature $L^{\prime}$ photometry, it was instead estimated using the $W1$ magnitude from the AllWISE catalog \citep{Cutri:2014wx}, and the empirical spectral type vs. $L^{\prime}-W1$ color relation given in Table~\ref{tab:filters}. As the spectra within the SpeX library were not flux calibrated, they were scaled to best fit the 2MASS $J$, $H$, and $K_{\rm S}$ apparent magnitudes for each object.

\begin{figure}
\epsscale{1.2}
\plotone{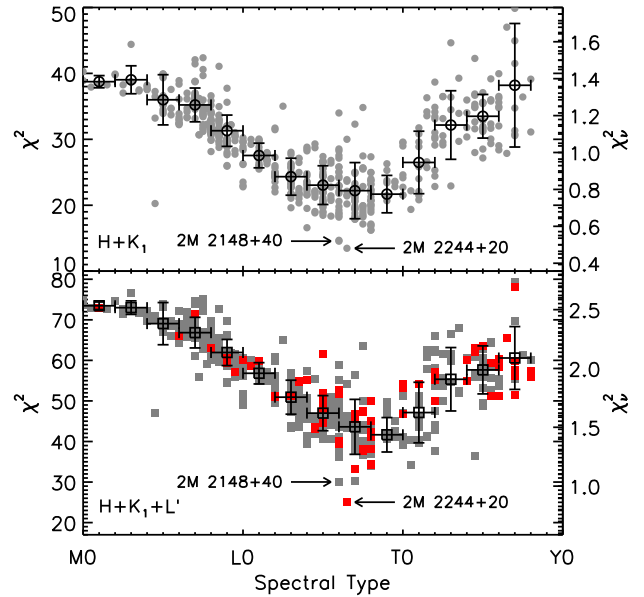}
\caption{The minimum $\chi^2$ for each object within the SpeX Prism Spectral Library and the IRTF Spectral Library considering only the near-infrared SED ($H$ \& $K_1$, top panel) and full SED ($H$, $K_1$ \& $L^{\prime}$, bottom panel) of HD~95086~b. Those objects with literature MKO $L^{\prime}$ photometry are highlighted as filled red squares. The $L^{\prime}$ photometry for the remaining objects was estimated from their {\it WISE} $W1$ magnitudes, and the empirical $L^{\prime}-W1$ color transformation given in Table~\ref{tab:filters}. For both families of fits, the average and standard deviation of the $\chi^2$ values were computed in bins two spectral subtypes wide (open symbols). From these fits, the spectral type of HD~95086~b was found to be between L$7\pm6$ and L$9_{-6}^{+4}$ using the near-infrared and full SEDs, respectively. The lower and upper bounds are defined as the spectral type bin at which the average of the $\chi^2$ of that bin is significantly higher than that of the minimum average $\chi^2$. Two of the best fitting field brown dwarfs, 2M~2244+20 and 2M~2148+40, are indicated.}
\label{fig:spex}
\end{figure}
For each comparison object, two fits were attempted: a fit to only the $H$-band photometry and $K_1$ spectrum (Figure~\ref{fig:spex}, lower curve), and a fit also including the $L^{\prime}$ photometry (Figure~\ref{fig:spex}, upper curve). Synthetic $H$-band photometry was calculated for each object using the procedure described in the Appendix, and a synthetic $K_1$ spectrum at the resolution of GPI was created by degrading the resolution of the library spectrum by convolution with a Gaussian kernel. The goodness of fit $\chi^2$ was then calculated for each comparison object as
\begin{equation}
    \chi^2 = R^{\rm T}C^{-1}R + \sum_i \left(\frac{f_i - \alpha s_i}{\sigma_i}\right)^2.
\end{equation}

The first term computes the $\chi^2$ of the fit to the $K_1$ spectrum alone, accounting for the covariances between neighbouring wavelength channels. $R = F-\alpha S$, the residual vector, is the difference between the $K_1$ spectrum of HD~95086~b $F = \{f_1\cdots f_n\}$, and the $K_1$ spectrum of the comparison object $S = \{s_1\cdots s_n\}$ multiplied by a scaling factor $\alpha$, where $n=29$ the number of wavelength channels within the $K_1$ spectrum. $C$ is the covariance matrix calculated from the parametrized noise model described in Section~\ref{sec:covariance}. The second term incorporates the $H$ and $L^{\prime}$ photometry of HD~95086~b in the standard $\chi^2$ formalism: $f_i$ and $\sigma_i$ are the flux and corresponding uncertainty of HD~95086~b in the i$^{\rm th}$ filter, $\alpha$ is the same multiplicative factor as in the first term, and $s_i$ is the flux of the comparison object in the same filter. For each comparison object, the value of the multiplicative factor $\alpha$ which minimizes $\chi^2$ was found using a truncated-Newton minimization algorithm. The resulting minimum $\chi^2$ are plotted as a function of spectral type in Figure~\ref{fig:spex}.

Considering only the fit to the near-infrared SED of HD~95086~b, $\chi^2$ reaches a minimum at spectral type of L7, although this minimum is poorly constrained between L1 and T3 (Fig.~\ref{fig:spex}, top panel). Later T-dwarfs are strongly excluded due to the decline in flux seen in the $K$-band beyond 2.08~\micron~caused by increasing CH$_4$ and H$_2$O absorption within the atmospheres of these cool objects \citep{Kirkpatrick:2005cv}, compared with the monotonically rising spectrum of HD~95086~b. The addition of the $L^{\prime}$ photometry to the fit results in a significant increase in the value of $\chi^2$ at all spectral types (Fig.~\ref{fig:spex}, bottom panel), consistent with the unusual location of HD~95086~b on the CCD and CMDs (Figures~\ref{fig:CCD}, \ref{fig:CMDa}, \ref{fig:CMDb}). Fitting to the full SED of HD~95086~b shifted the minimum $\chi^2$ to the later spectral type of L9, with the minimum similarly poorly constrained between L3 and T3.

The best fitting object was the very red L6.5 brown dwarf 2M~2244+20 \citep{Dahn:2002fu}, which possesses an extremely cloudy atmosphere \citep{Leggett:2007if}, may have a low surface gravity \citep{McLean:2003hx,Looper:2008hs}, and was found to be the best match to the near-infrared SED of HR~8799~b \citep{Bowler:2010ft}. Another good fit was the L6 brown dwarf  2M~2148+40 \citep{Looper:2008hs}, which has been identified as being old, but with extremely red near-infrared colors indicative of thick clouds \citep{Cruz:2009gs}. The spectral type of HD~95086~b was also estimated using the empirical absolute magnitude-spectral type relations given in \citet{Dupuy:2012bp} for field brown dwarfs. Using the absolute magnitudes of HD~95086~b given in Table~\ref{tab:properties}, the spectral type was estimated as T$7\pm0.5$, T$3\pm1.5$, and L$9\pm2$ using $M_H$, $M_{K_1}$, and $M_{L^{\prime}}$, respectively. These estimates further emphasize the peculiar morphology of the SED of HD~95086~b, and how discrepant it is from that of old field brown dwarfs.

\subsubsection{Comparison to young substellar companions and isolated young/dusty brown dwarfs}
\label{sec:comparison_to_young}
\begin{figure}
\epsscale{1.2}
\plotone{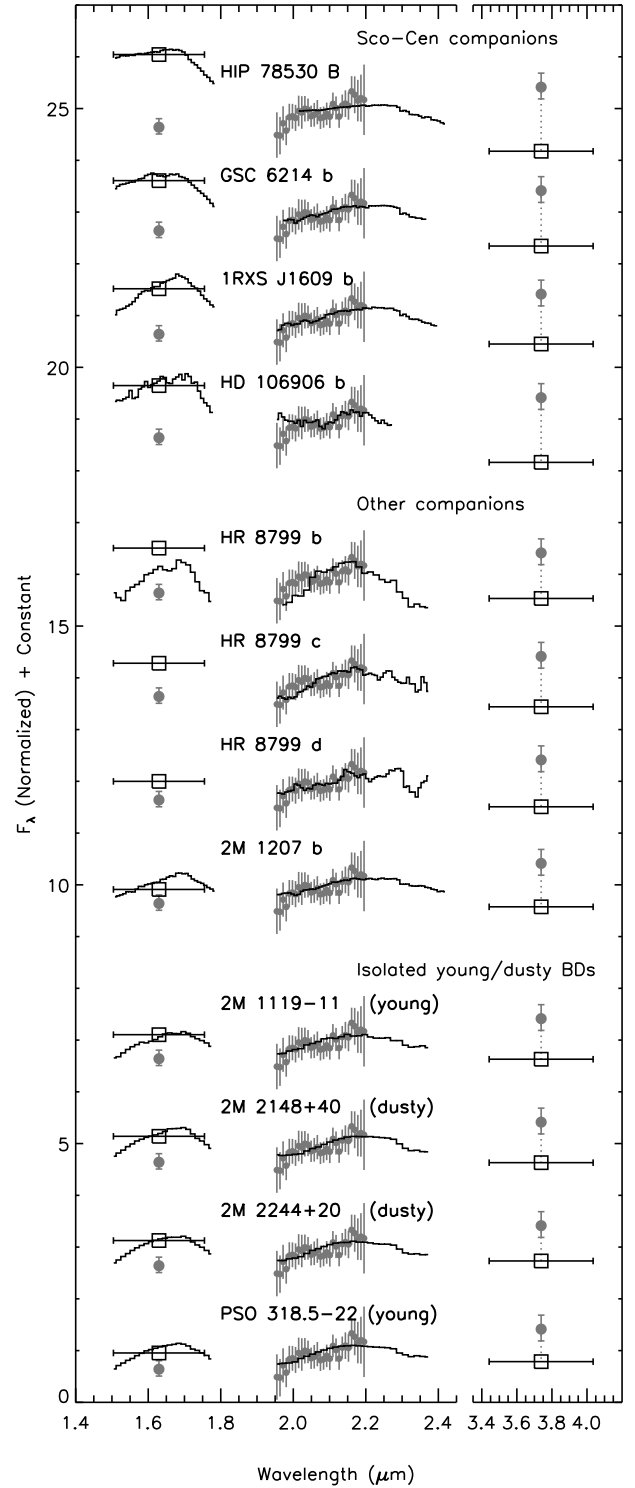}
\caption{Spectral energy distribution of HD~95086~b (gray points) compared with several substellar companions within Sco-Cen (top four), other young planetary-mass companions (middle four), and isolated young and/or dusty brown dwarfs (bottom four). The spectral energy distribution of each object has been normalized by the flux within the $K_1$ bandpass. For clarity, dotted lines connect the $L^{\prime}$ flux of HD~95086~b and the comparison object. The discrepancy between the $H$-band photometry and $H$-band spectrum for HR~8799~b is because the $H$-band spectrum was normalized by \citet{Barman:2011fe} using $H=18.06$ \citep{Metchev:2009jl}, whereas the plotted $H$-band photometry uses $H=17.88$ \citep{Esposito:2013hs}.}
\label{fig:comparison}
\end{figure}
The complete SED of HD~95086~b was also compared to that of substellar companions within Sco-Cen, and around other young stars, and of isolated young and/or dusty brown dwarfs in Figure~\ref{fig:comparison}. For each comparison object, the SED was normalized to the integrated flux in the $K_1$ filter bandpass. The $K_1$ spectrum of HIP~78530~B \citep{Lafreniere:2011dh} and HIP~106906~b \citep{Bailey:2014et} are flatter than that of HD~95086~b, which may indicate a lower effective temperature and a spectral type later than L2.5 for HD~95086~b. The $K_1$ spectrum of HD~95086~b is most similar to that of GSC~6214~b \citep{Bowler:2011gw}, 1RXS~1609~b \citep{Lafreniere:2010cp}, 2M~1207~b \citep{Patience:2010hf}, and HR~8799~d \citep{Ingraham:2014gx}. HR~8799~b \citep {Barman:2011fe} and c \citep{Ingraham:2014gx} have a somewhat steeper slope in $K_1$ than HD~95086~b. The four young and/or dusty brown dwarfs have very similar $K_1$ spectra, consistent with their similar effective temperatures and spectral types. These comparison objects span a range of effective temperatures (750--2700~K), and spectral types (from mid-M to L/T transition objects). With the exception of HIP~78530~B, which exhibits an almost flat $K_1$ spectrum due to its high effective temperature (2700~K), there is very little diversity in the shape of the $K_1$ spectra, despite the large range of effective temperatures. This suggests that the shape of the $K_1$ spectrum on its own is a poor diagnostic of the effective temperatures of young planetary-mass objects when at a resolution and SNR similar to the HD~95086~b spectrum.

\section{Model Atmosphere Fitting}
\label{sec:models}
\begin{deluxetable*}{lcccccccccc}
\tablecaption{Model grid properties and best fit}
\tablewidth{0pt}
\tablehead{
\colhead{Model} & \multicolumn{4}{c}{Parameter Range/Spacing} & \multicolumn{4}{c}{HD 95086 b}\\
& $T_{\rm eff}$ & $\Delta T_{\rm eff}$& $\log g$ & $\Delta\log g$ & $T_{\rm eff}$ & $\log g$ & $R$ & $\log L/L_{\odot}$ & $\chi^2$ & $\chi^2_{\nu}$\\
&(K) & (K) & (dex) & (dex) & (K) & (dex) & ($R_{\rm Jup}$) & (dex) & }
\startdata
\multicolumn{10}{c}{Cloudy/dusty Atmospheres}\\
\noalign{\smallskip}\hline\noalign{\smallskip}
{\sc AMES-Dusty} & \phantom{0}500--1500 & 100 & 3.50--6.00 & 0.50 & 1200 & 5.00 & 0.83 & -4.86 & 14.35 & 0.55\\
&& 5 && 0.01 & 1250 & 3.50 & 0.79 & -- & 13.51 & 0.52\\
{\sc BT-Dusty} & 1000--1500 & 100 & 4.50--5.50 & 0.50 & 1000 & 4.50 & 1.24 & -4.86 & 18.42 & 0.71\\
&& 5 && 0.01 & 1000 & 4.50 & 1.24 & -- & 18.42 & 0.71\\
{\sc Drift-Phoenix} & 1000--1500 & 100 & 3.00--6.00 & 0.50 & 1300 & 3.50 & 0.72 & -4.88 & 11.53 & 0.44\\
&& 5 && 0.01 & 1270 & 3.50 & 0.77 & -- & 10.99 & 0.42\\
Madhusudhan (A60-F) & \phantom{0}700--1500 & 100 & 3.75--4.25 & 0.25 & 800 & 3.75 & 1.87 & -4.90 & 16.26 & 0.63\\
&& 5 && 0.01 & 780 & 3.75 & 2.00 & -- & 15.31 & 0.59\\
\noalign{\smallskip}\hline\noalign{\smallskip}
\multicolumn{10}{c}{Clear Atmospheres}\\
\noalign{\smallskip}\hline\noalign{\smallskip}
{\sc AMES-Cond} & \phantom{0}500--1500 & 100 & 2.50--6.00 & 0.50 & 1000 & 2.50 & 0.81 & -5.23 & 47.92 & 1.84\\
&& 5 && 0.01 & 965 & 2.68 & 0.88 & -- & 47.82 & 1.84\\
{\sc BT-Cond} & \phantom{0}800--1500 & 100 & 4.00-5.50 & 0.50 & 1000 & 4.00 & 1.00 & -5.05 & 49.95 & 1.92\\
&& 5 && 0.01 & 975 & 4.00 & 1.07 & -- & 49.93 & 1.92\\
{\sc BT-Settl} & \phantom{0}500--1500 & 50 & 3.00--5.50 & 0.50 & 1150 & 3.50 & 0.69 & -5.17 & 20.80 & 0.80\\
&& 5 && 0.01 & 1150 & 3.51 & 0.68 & -- & 20.80 & 0.80
\enddata
\label{tab:models}
\end{deluxetable*}

Combining the revised $H$-band photometry and the new $K_1$ spectrum obtained with GPI with literature $L^{\prime}$ photometry (Table \ref{tab:properties}; \citealp{Galicher:2014er}) further constrains the near-infrared SED of HD~95086~b. The SED was fit to seven publicly available grids of model atmospheres: {\sc AMES-Cond}\footnote{{\tt https://phoenix.ens-lyon.fr/Grids/AMES-Cond/SPECTRA/}} \citep{Allard:2001fh,Baraffe:2003bj}, {\sc AMES-Dusty}\footnote{{\tt https://phoenix.ens-lyon.fr/Grids/AMES-Dusty/SPECTRA/}} \citep{Chabrier:2000hq, Allard:2001fh}, {\sc BT-Cond}\footnote{\tt https://phoenix.ens-lyon.fr/Grids/BT-Cond/}, {\sc BT-Dusty}\footnote{\tt https://phoenix.ens-lyon.fr/Grids/BT-Dusty/AGSS2009} and {\sc BT-Settl}\footnote{ \tt https://phoenix.ens-lyon.fr/Grids/BT-Settl/CIFIST2011} (e.g., \citealp{Allard:2012fp}), {\sc Drift-Phoenix}\footnote{\tt http://svo2.cab.inta-csic.es/theory/main} (e.g., \citealp{Helling:2008ht}), and the various models published by A. Burrows and collaborators\footnote{\tt http://www.astro.princeton.edu/\char`\~burrows} \citep{Burrows:2006ia,Hubeny:2007hm,Madhusudhan:2011ex}. The range of effective temperatures ($T_{\rm eff}$) and surface gravities ($\log g$), and the grid spacing for each parameter, is given for each model grid in Table~\ref{tab:models}. Due to the irregular sampling of some grids at non-solar metallicities, a solar metallicity for HD~95086~b was assumed.

\subsection{{\sc AMES-Cond} \& \sc{AMES-Dusty}}
The AMES family of models \citep{Allard:2001fh} were created by combining the {\tt PHOENIX} plane-parallel radiative transfer atmosphere model \citep{Hauschildt:1992ff} with the NASA AMES molecular H$_2$O and TiO line lists \citep{Partridge:1997kh}, using the solar abundances of \citet{Grevesse:1993uj}. The {\sc AMES-Dusty} grid represents a limiting case including treatments for both the condensation of dust within the photosphere, and for dust opacities. Dust grain sizes were drawn from the interstellar grain size distribution \citep{Mathis:1977hp}, and opacities were calculated assuming Mie scattering for spherical grains. Gravitational sedimentation was ignored, leading to a significant amount of photospheric dust. The effect of the dust was to increase the gas temperature, which in turn decreased the strength of molecular absorption features in the optical and near-infrared. On the other extreme, the {\sc AMES-Cond} grid included the treatment for dust condensation, but ignored dust opacities entirely in order to simulate the immediate sedimentation of dust to the lower atmosphere, leading to a dust-free photosphere \citep{Allard:2001fh}.

\subsection{{\sc BT-Cond}, {\sc BT-Dusty} \& {\sc BT-Settl}}
As with the AMES models, the BT family of models (e.g., \citealp{Allard:2012fp}) also used the {\tt PHOENIX} atmosphere model, but with updated molecular line lists, most notably for H$_2$O \citep{Barber:2006dm}, and revised solar abundances of \citet{Asplund:2009eu} for {\sc BT-Dusty} and \citet{Caffau:2011ik} for {\sc BT-Cond} and {\sc BT-Settl}. The {\sc BT-Cond} and {\sc BT-Dusty} grids used the same treatment for dust as in their counterpart within the {\sc AMES} grid, while the {\sc BT-Settl} grid used a more detailed cloud model to define the number density and size distribution of condensates within the atmosphere \citep{Allard:2012fp}. The {\sc BT-Settl} models are able to reproduce the L/T transition observed for field brown dwarfs, where the atmosphere transitions from fully cloudy to fully clear (e.g., \citealp{Marley:2002il}), due in part to the formation of methane at lower temperatures.

\subsection{{\sc Drift-Phoenix}}
The {\sc Drift-Phoenix} grids \citep{Woitke:2003cs,Woitke:2004ie,Helling:2006gp,Helling:2008ht} also used the {\tt PHOENIX} model with the same solar abundances as the AMES models but is coupled with the non-equilibrium cloud model {\tt DRIFT} which includes treatments for the formation, growth, evaporation, settling, and advection of grains \citep{Helling:2008ht}. Contrary to the other models, they used a kinetic approach to describe the formation of grains, and a top-down approach to simulate the motion of the grains within the atmosphere. The {\sc Drift-Phoenix} models have been successful in reproducing the near-infrared SEDs of young substellar companions over a wide range of effective temperatures (e.g. \citealp{Patience:2012cx,Bonnefoy:2014dh,Lachapelle:2015cx}).

\subsection{Madhusudhan et al.}
The models of A. Burrows and collaborators \citep{Burrows:2006ia, Hubeny:2007hm, Madhusudhan:2011ex} use a variant of the {\tt TLUSTY} model stellar atmosphere code \citep{Hubeny:1988ci,Hubeny:1995fw} where the opacity as a function of wavelength is pre-computed using a chemical equilibrium code to account for the sedimentation of particles in a gravitational field \citep{Hubeny:2003eb}. The \citet{Burrows:2006ia} models are able to reproduce the overall shape of the L/T transition, although no single combination of surface gravity and cloud particle size can fit all of the observations. Using an updated version of the atmospheric model, \citet{Hubeny:2007hm} explored the effects of non-equilibrium chemistry, although these effects are primarily limited to the $M$ ($4.8$~$\micron$) and $N$ ($10.5$~$\micron$) bands (e.g., \citealp{Saumon:2000bb}).

Although these models were able to reproduce the overall shape of the L through T sequence, they were a poor fit to the observations of the HR~8799 planets. By invoking a significantly thicker cloud layer and adjusting the modal dust particle size, \citet{Madhusudhan:2011ex} were able to reproduce the available photometry of the three outermost planets of the HR~8799 system, which are displaced from field brown dwarf sequence on the CMD. \citet{Madhusudhan:2011ex} considered four models of the cloud structure, modal particle sizes between 1--100~\micron, and spherical dust grains composed of either fosterite or iron. While the SED of HD~95086~b was fit to each of the nine grids from \citet{Madhusudhan:2011ex}, and the 21 grids from \citet{Burrows:2006ia} and \citet{Hubeny:2007hm}, only the results of the grid containing the best fit are discussed below (Model~A, 60~\micron~modal particle size, fosterite grains; \citealp{Madhusudhan:2011ex}).

\subsection{Fitting Procedure}
For the individual spectra within each model grid the fitting procedure was the same. Synthetic $H$ and $L^{\prime}$ photometry was computed for each model spectrum by folding the spectrum through the GPI $H$ and NaCo $L^{\prime}$ filter transmission curves. Synthetic $K_1$ spectra were also computed by degrading the resolution of the model spectra to that of the GPI $K1$ data ($\lambda/\delta\lambda \approx 66$), and interpolating the resulting smoothed spectra to the same wavelength values. The goodness of fit ($\chi^2$) for each model to the measurements of HD~95086 was calculated using the method described in Section~\ref{sec:spex_comparison}, where $\alpha = R^2/d^2$ was allowed to vary such that the radius $R$ was in the range 0.6--2.0 $R_{\rm Jup}$ as in \citet{Galicher:2014er}, given the distance $d$ to HD~95086 measured by {\it Hipparcos} \citep{vanLeeuwen:2007dc}. This process was repeated on a finer version of the grid, obtained by linearly interpolating between the logarithm of the flux at each grid point, where the spacing in $T_{\rm eff}$ and $\log g$ were reduced to 5~K and 0.01~dex in order to evaluate the confidence interval on each parameter. The 68\%, 95\%, and 99\% confidence regions were calculated by integrating the likelihood ${\mathcal L}$ over the full parameter range, where $\mathcal L = {\rm e}^{-\chi^2/2}$.

\subsection{Effective temperature and surface gravity}
\label{sec:fitted}
\begin{figure*}
\epsscale{1.2}
\plotone{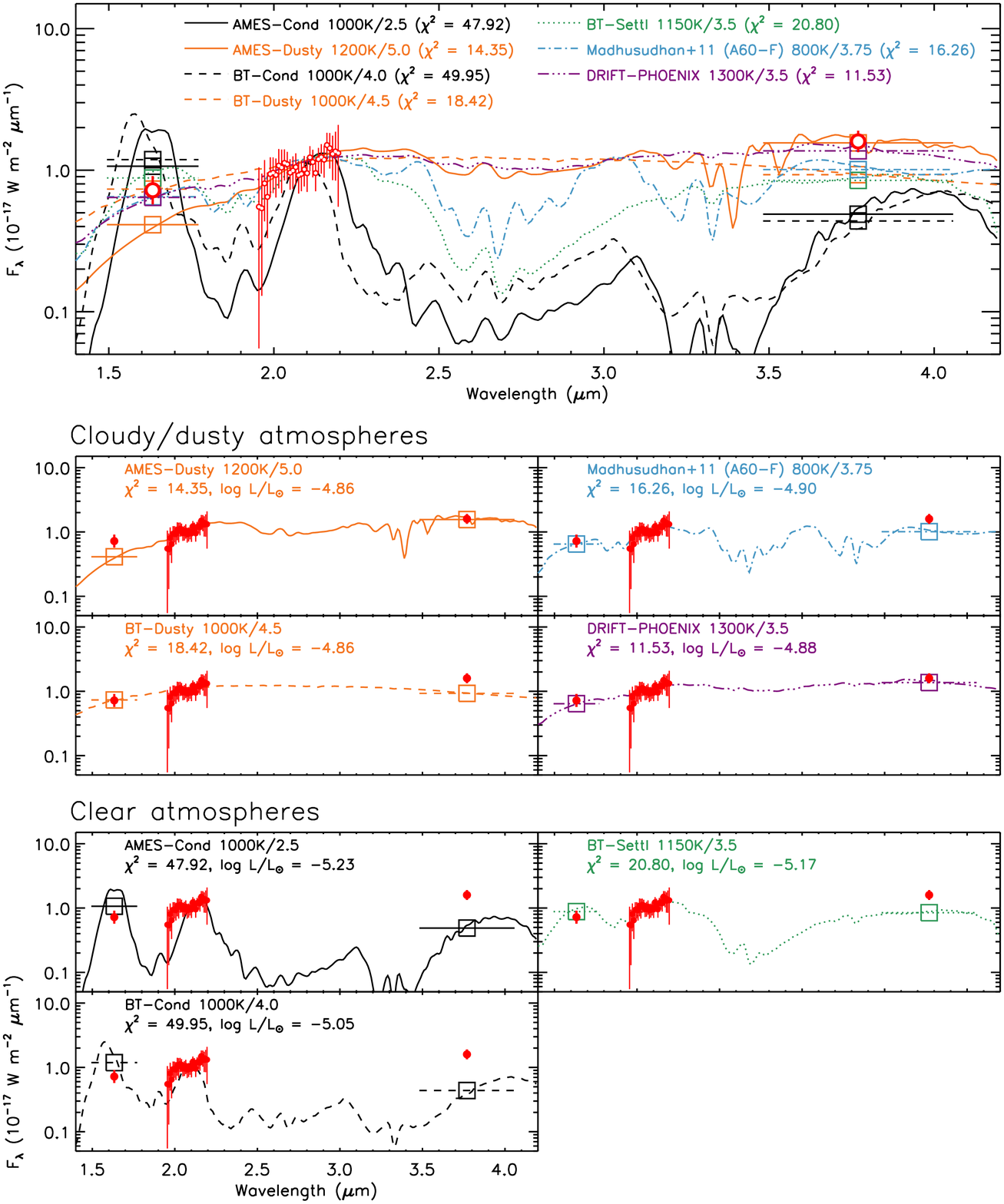}
\caption{Photometry and spectroscopy of HD~95086~b (red open circles) and best-fit model atmosphere spectra. Synthetic $H$ and $L^{\prime}$ photometry for the model spectra are plotted as open squares, with the vertical error bar corresponding to the effective width of the filter. The top panel contains the best fit from each model grid for ease of comparison, while the bottom panels contain each fit individually.}
\label{fig:best_fit}
\end{figure*}
\begin{figure*}
\epsscale{1.2}
\plotone{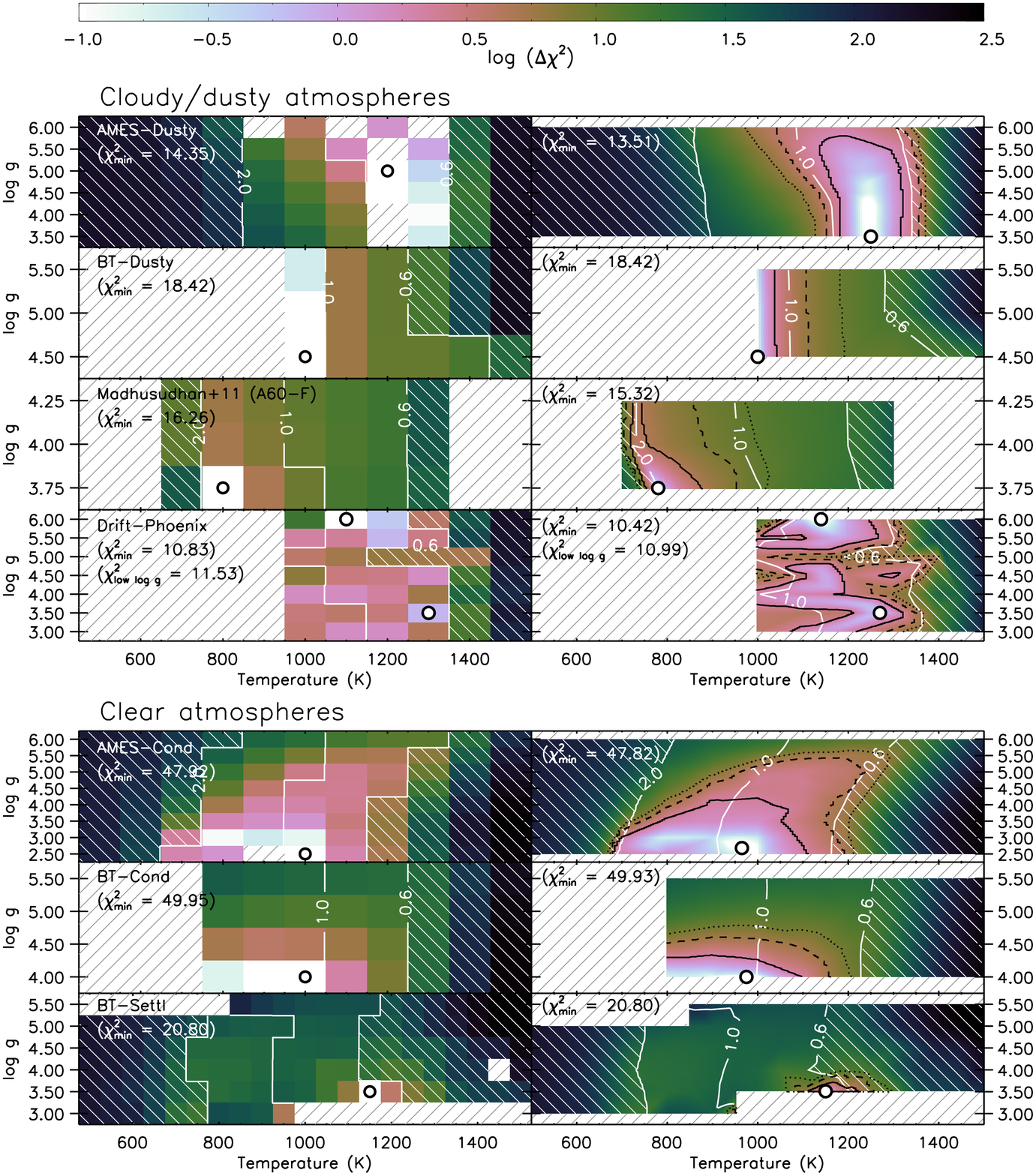}
\caption{$\Delta \chi^2 \equiv \chi^2 - \chi^2_{\rm min}$ surfaces for the seven model grids (left column), and the six interpolated grids (right column) plotted as a function of effective temperature and surface gravity. For each grid, the best fitting model is indicated by the circle symbol. For the {\sc Drift-Phoenix} grid, the best fit low surface gravity model is also indicated. For the interpolated grids, the black contours indicate the 68\% (solid), 95\% (dashed), and 99\% (dotted) confidence regions. White contours define radii of 0.6, 1.0, and 2.0~$R_{\rm Jup}$, based on the dilution factor required to minimize $\chi^2$ for each spectrum. Positive hatching corresponds to regions of phase space within each grid for which no model spectra exist. Negative hatching indicate models with a dilution factor corresponding to a radius of $R<0.6$~$R_{\rm Jup}$ and $R>2.0$~$R_{\rm Jup}$, which are not physical for HD~95086~b \citep{Mordasini:2012gp}. Only the results of the fit to the spectra within the Model~A, 60~\micron~modal grain size, fosterite grain grid are shown for the \citet{Madhusudhan:2011ex} grid.}
\label{fig:fit_surface}
\end{figure*}
The effective temperature and surface gravity of the best fit spectrum from each of the seven model grids, and their interpolated versions, are given in Table~\ref{tab:models}. For each fit, the reduced $\chi^2$ was calculated assuming 28 degrees of freedom. The best-fit spectra within the original non-interpolated grids are plotted in Figure~\ref{fig:best_fit}. The $\Delta \chi^2 \equiv \chi^2 - \chi^2_{\rm min}$ surfaces for each grid are plotted as a function of temperature and surface gravity in Figure~\ref{fig:fit_surface}. Of the \citet{Madhusudhan:2011ex} models, the best fit was found in the Model~A grid, the thickest (in vertical extent) of the four cloud structures simulated, and thicker than the clouds required to explain the SEDs of the HR~8799 planets \citep{Madhusudhan:2011ex}. As such, only the $\Delta\chi^2$ surface from this particular grid is shown in Figure~\ref{fig:fit_surface}.

The SED of HD~95086~b was best fit by an atmosphere within the {\sc Drift-Phoenix} grid which simulates a significant photospheric dust content. Within this grid, two distinct minima in the $\Delta \chi^2$ surface were found (Fig.~\ref{fig:fit_surface}). One was found at high surface gravities ($\log g = 6.0$) at a temperature of 1100~K, with a $\chi^2=10.83$ and a radius of 0.95~$R_{\rm Jup}$; and the other at a lower surface gravity ($\log g = 3.5$) at 1300~K, with a $\chi^2=11.53$ and a smaller radius of 0.72~$R_{\rm Jup}$. As young planetary-mass objects are expected to have relatively low surface gravities, only the lower surface gravity fit is reported in Table~\ref{tab:models}. The SED was also fit well by an {\sc AMES-Dusty}  $T_{\rm eff} = 1200$~K, $\log g = 5.0$ model atmosphere, with a radius of 0.83~$R_{\rm Jup}$ ($\chi^2 = 14.35$).  Unfortunately, the {\sc AMES-Dusty} grid\footnote{{\tt https://phoenix.ens-lyon.fr/Grids/AMES-Dusty/SPECTRA/}} contains several erroneous models at 1200~K: the $\log g = 5.5$ model is incomplete, and the $\log g = $ 3.5--4.5 models are all identical. For the purposes of this study, we assume that these three identical models have $\log g = 4.5$. In the interpolated grid, where these gaps are interpolated over, the best fit is at a similar $T_{\rm eff}=1250$~K, but at a lower $\log g = 3.50$ ($\chi^2 = 13.51$). The two remaining grids which simulate a dusty atmosphere---the Madhusudhan et al. and {\sc BT-Dusty} models---both predict a low surface gravity, although the goodness of fit statistic for each are worse at $\chi^2 = 16.26$ and $\chi^2 = 18.42$. In each case the best fit model is railed at the grid boundary, suggesting that an expanded parameter search is necessary to locate the true minimum $\chi^2$.

The fits to the clear atmospheres are worse, with a goodness of fit ranging between $\chi^2=20.80$ for the {\sc BT-Settl} grid and $\chi^2=49.95$ for the {\sc BT-Cond} grid. The best-fitting temperatures for HD~95086~b within these grids are cooler than the predicted temperature of the L-T transition, where brown dwarfs transition from having dusty to clear atmospheres with decreasing temperature. The position of HD~95086~b on the CMD (Figures~\ref{fig:CMDa} and \ref{fig:CMDb}) in the extension to the L-dwarf sequence, is inconsistent with the predicted luminosity and colors of objects with fully clear atmospheres. As the current observational evidence is consistent with a dusty atmosphere, we use only the fits to the dusty atmosphere models to estimate a $T_{\rm iff}$ and $\log g$ for HD~95086~b. Unfortunately, the variation in which part of parameter space is sampled by each grid means that these two parameters are still relatively unconstrained. In terms of $T_{\rm iff}$, best fit models were found between 800--1300~K, whereas $\log g$ can only be constrained to be $\lesssim 4.5$ as the best fit is railed at the lowest surface gravity within three of the four model grids.

\subsection{Bolometric Luminosity and Mass}
\begin{figure}
\epsscale{1.2}
\plotone{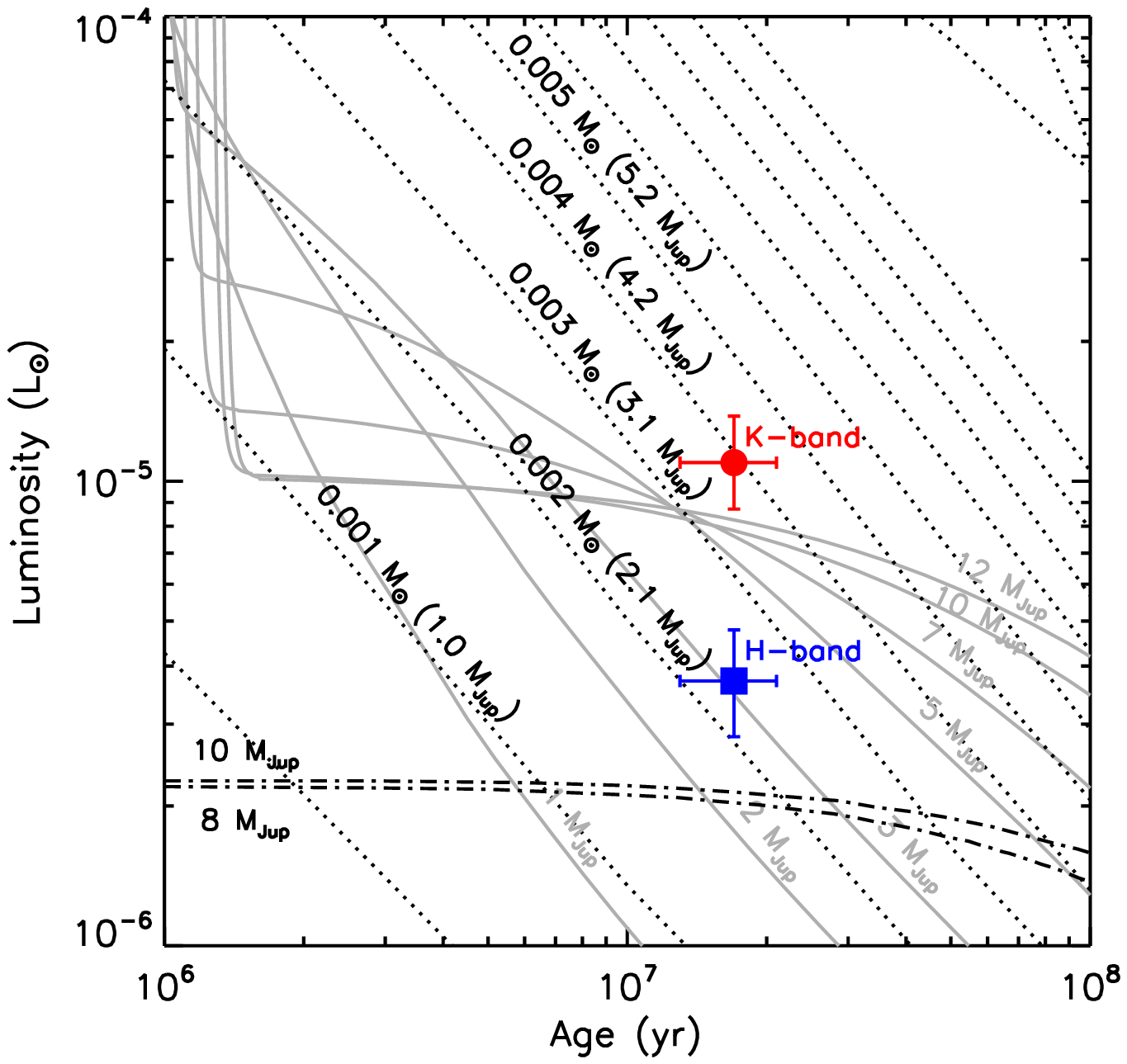}
\caption{Luminosity as a function of age for substellar objects from the ``hot-start'' evolutionary models of  \citet{Baraffe:2003bj} (dotted curves), and the ``cold-start'' evolutionary models of \citet{Marley:2007bf} (dot-dashed curves) and \citet{Mordasini:2013cr} (solid gray curves). For clarity, only the two most massive models from the \citet{Marley:2007bf} grid are plotted due to the degeneracy between mass and luminosity at the age of HD~95086~b. Two estimates of the luminosity of HD~95086~b, derived from the two bolometric corrections, are plotted (blue square: $H$-band, red circle: $K$-band).}
\label{fig:luminosity}
\end{figure}
\begin{deluxetable}{lcccc}
\tablecaption{Physical properties of HD~95086~b estimated from the bolometric luminosity}
\tablewidth{0pt}
\tablehead{
\colhead{Model} & \colhead{Band} & \colhead{$M$} & \colhead{$R$} & \colhead{$T_{\rm eff}$}\\
\colhead{}      &\colhead{}      & \colhead{($M_{\rm Jup}$)} & \colhead{($R_{\rm Jup}$)} & \colhead{(K)}}
\startdata
Baraffe   & $H$ & $2.7^{+0.6}_{-0.5}$ & $1.32^{+0.02}_{-0.01}$ & $703^{+47}_{-43}$\\
          & $K$ & $4.4^{+0.8}_{-0.8}$ & $1.34^{+0.02}_{-0.02}$ & $910^{+54}_{-47}$\\
Mordasini & $H$ & $3.1^{+1.0}_{-0.7}$ & $1.30^{+0.01}_{-0.01}$ & $702_{-46}^{+51}$\\
          & $K$ & $>9.1$ & - & -
\enddata
\label{tab:bolometric}
\end{deluxetable}

Using the revised $H$ and $K_1$ magnitudes presented in this study, the bolometric luminosity of HD~95086~b was estimated. The GPI magnitudes were converted into MKO $H$ and $K$ magnitudes using the empirical filter transformations in Table~\ref{tab:filters}, assuming a spectral type of L$7\pm6$ (\S \ref{sec:spectrum}). The bolometric corrections for each band were then estimated from the spectral type as $BC(H)=2.53^{+0.13}_{-0.08}$ from the fit to field objects of \citet{Liu:2010cw}, and $BC(K)=3.19^{+0.13}_{-0.38}$ from the fit to young low-surface gravity objects of \citet{Filippazzo:2015dv}, corresponding to bolometric magnitudes of $M_{\rm bol}(H)=18.30\pm0.28$ and $M_{\rm bol}(K)=17.15^{+0.24}_{-0.26}$. Assuming a solar bolometric magnitude of $M_{\rm bol, \odot}=4.74$,  these bolometric magnitudes correspond to luminosities of $\log L/L_{\odot} =-5.42\pm0.11$ using the $H$-band correction, and $\log L/L_{\odot} = -4.96 \pm 0.10$ using the $K$-band correction. While the $K$-band bolometric correction was based on young low-surface gravity objects, the $H$-band correction was based on field brown dwarfs, which have significantly different SEDs to that of young substellar objects. For instance, \citet{Filippazzo:2015dv} measured a one magnitude offset in the value of $BC(J)$ between field and young objects. Given the unusual nature of the SED of HD~95086~b, the bolometric magnitudes, and the associated derived luminosities, may be similarly biased. The bolometric luminosity was also estimated by integrating the best fit model atmosphere from each of the model grids, ranging from $\log L/L_{\odot}=-5.23$ to $-4.86$ (Table \ref{tab:models}), with cloudy atmosphere models predicting higher luminosities ($-4.90$ to $-4.86$) than clear atmosphere models ($-5.23$ to $-5.05$).

As the age of HD~95086~b ($17\pm4$~Myr) is well constrained due to membership of the Lower Centaurus Crux subgroup of the Sco-Cen association, the luminosity can be used to obtain a model-dependent estimate of the mass. The luminosity of HD~95086~b is plotted relative to three evolutionary models in Figure~\ref{fig:luminosity}. At the two extremes are the ``hot-start'' \citet{Baraffe:2003bj} models, in which gas accreting onto the forming planet is unable to efficiently cool resulting in a high initial luminosity and entropy, and the ``cold-start'' models of \citet{Marley:2007bf}, in which the accreting gas has radiated away most of its heat, resulting in a low initial luminosity and entropy for the planet. Intermediate to these are the models of \citet{Mordasini:2013cr}, which predict a strong positive correlation between the luminosity and the mass of the solid core of the planet. In order to assess the range of probable masses for HD~95086~b, only the evolutionary models with the smallest core mass of 22~M$_{\oplus}$ were used from the \citet{Mordasini:2013cr} grid. Higher core masses result in a higher initial luminosity for a given final planet mass, converging with the luminosity predictions of the \citet{Baraffe:2003bj} models for core masses $>100$~M$_{\oplus}$.

The mass, radius, and effective temperature derived from fitting both the $H$- and $K$-band luminosity to the \citet{Baraffe:2003bj} and \citet{Mordasini:2013cr} evolutionary models are listed in Table~\ref{tab:bolometric}. No attempt was made to use the \citet{Marley:2007bf} models as the two luminosity estimates were significantly higher than that for the most massive object within the grid, as shown in Figure~\ref{fig:luminosity}. Based on the \citet{Baraffe:2003bj} models, the mass of HD~95086~b was estimated to be between $2.7^{+0.6}_{-0.5}~M_{\rm Jup}$, using the $H$-band luminosity, and $4.4\pm0.8$~$M_{\rm Jup}$, using the $K$-band luminosity. The mass estimated from the \citet{Mordasini:2013cr} models using the $H$-band luminosity ($3.1^{+1.0}_{-0.7}~M_{\rm Jup}$) was consistent with the two previous estimates, however as the $K$-band luminosity was significantly higher than that predicted for the most massive object within the grid, only a lower-limit on the mass of $M>9.1$~$M_{\rm Jup}$ could be estimated. The quoted uncertainties on the model-dependent masses incorporate the uncertainty on the age of HD~95086~b, and the uncertainty on the estimated bolometric luminosity derived from the $H$- and $K$-band bolometric corrections.

\section{Concluding Remarks}
We have presented the first spectroscopic measurement of the young, low-mass, and extremely red exoplanet HD~95086~b. Obtained with the Gemini Planet Imager and spanning the blue half of the $K$-band ($K_1$, 1.9--2.2~\micron), the spectrum was recovered at an SNR of 3--5 within the individual wavelength channels, and shows a monotonic increase in flux towards longer wavelengths within the measurement uncertainties. Combining the new $K_1$ spectrum and the revised $H$ photometry with literature NaCo $L^{\prime}$ photometry \citep{Galicher:2014er}, the SED of the planet between 1.5--4.0~\micron~was constructed and compared with other young substellar objects, and older field brown dwarfs. When placed on the MKO $M_{L^{\prime}}$ vs. GPI $K_1$ $-$ MKO $L^{\prime}$ color-magnitude diagram (Figure~\ref{fig:CMDb}), HD~95086~b occupies a region of the diagram devoid of any comparable objects. While similar in $L^{\prime}$ luminosity to field late L-dwarfs, HD~95086~b is over a magnitude redder, and while having a similar $K_1-L^{\prime}$ color to field late T-dwarfs, it is over two magnitudes brighter. Comparing to young substellar objects, HD~95086~b is most analogous to 2M~1207~b, with a similar $L^{\prime}$ luminosity, but almost a magnitude redder in $K_1-L^{\prime}$. 2M~1207~b was itself noted as being extremely red \citep{Chauvin:2004cy}, which was explained by invoking thick clouds and non-equilibrium chemistry \citep{Barman:2011dq,Skemer:2014hy}. The unusually red color of HD~95086~b may also be explained by the presence of circumplanetary material, however further observations sensitive to emission from accretion are require to test this hypothesis.

Comparing the near-infrared portion of the SED of HD~95086~b to field brown dwarfs within the SpeX Prism and IRTF Spectral Libraries, a best fit was found at L7, although the relatively sparse coverage of the SED for HD~95086~b led to a broad minimum spanning between L1 and T3 (Figure~\ref{fig:spex}). The best fitting objects within the library were the brown dwarfs 2M~2244+20 and 2M~2148+40 which have been identified as possessing thick clouds, and in the case of 2M~2244+20, may have low surface gravity. The full SED of HD~95086~b was also compared with the predictions of grids of model atmospheres spanning a range of temperatures, surface gravities, and cloud properties (Figures~\ref{fig:best_fit} and \ref{fig:fit_surface}, and Table~\ref{tab:models}). Morphologically, the SED was best fit by the models which incorporate a high photospheric dust content, and was poorly fit by those with clear atmospheres. Considering the best fits within the dusty atmosphere grids, the effective temperature of HD~95086~b was constrained to between 800--1300~K, with a surface gravity of $\log g \lesssim 4.5$. A higher surface gravity field brown dwarf at the same effective temperature would have already transitioned to a clear photosphere, within which dust particulates would have rained out to the lower atmosphere. This places HD~95086~b in the extension to the L-dwarf sequence occupied by young low surface gravity substellar companions which have retained their cloudy atmospheres despite low effective temperatures (e.g., \citealp{Skemer:2012gr}).

HD~95086~b occupies a region of the color-magnitude diagram within which no comparable object exists, with an SED consistent with an atmosphere dominated by thick clouds. Future observations of HD~95086~b to perform near-infrared spectroscopy at $H$ and $K_2$ (covering the red half of the $K$-band atmospheric window), and to obtain thermal-infrared photometry at 3.3~\micron, will help distinguish between the predictions of cloudy and cloud-free models. Given its extremely red color, and the diagnostic power of thermal-infrared measurements, HD~95086~b also represents an ideal target for a future thermal-infrared integral field spectrograph in the Southern hemisphere, similar to those already being commissioned in the North \citep{Skemer:2015uf}.

\acknowledgements

The authors wish to thank Jonathan Gagn\'{e} for useful discussions regarding empirical comparisons, and the referee for their comments which helped improve the manuscript. This work was based on observations obtained at the Gemini Observatory, which is operated by the Association of Universities for Research in Astronomy, Inc., under a cooperative agreement with the National Science Foundation (NSF) on behalf of the Gemini partnership: the NSF (United States), the National Research Council (Canada), CONICYT (Chile), the Australian Research Council (Australia), Minist\'{e}rio da Ci\^{e}ncia, Tecnologia e Inova\c{c}\~{a}o (Brazil) and Ministerio de Ciencia, Tecnolog\'{i}a e Innovaci\'{o}n Productiva (Argentina). This research has made use of the SIMBAD database, operated at CDS, Strasbourg, France. This publication makes use of data products from the Two Micron All Sky Survey, which is a joint project of the University of Massachusetts and the Infrared Processing and Analysis Center/California Institute of Technology, funded by the National Aeronautics and Space Administration and the National Science Foundation. This publication makes use of data products from the Wide-field Infrared Survey Explorer, which is a joint project of the University of California, Los Angeles, and the Jet Propulsion Laboratory/California Institute of Technology, and NEOWISE, which is a project of the Jet Propulsion Laboratory/California Institute of Technology. WISE and NEOWISE are funded by the National Aeronautics and Space Administration. Supported by NSF grants AST-1518332 (R.J.D.R., J.R.G., J.J.W., P.K.), AST-1411868 (J.L.P., B.M., A.R.), and DGE-1311230 (K.W.D.). Supported by NASA grants NNX15AD95G/NEXSS and NNX15AC89G (R.J.D.R., J.R.G., J.J.W., P.K.), and NNX14AJ80G (B.M., E.L.N., F.M.). J.R., R.D. and D.L. acknowledge support from the Fonds de Recherche du Qu\'{e}bec. Portions of this work were performed under the auspices of the U.S. Department of Energy by Lawrence Livermore National Laboratory under Contract DE-AC52-07NA27344.

\appendix
\subsection{GPI filter characterization}
\label{sec:filter}
\begin{figure}
\epsscale{0.5}
\plotone{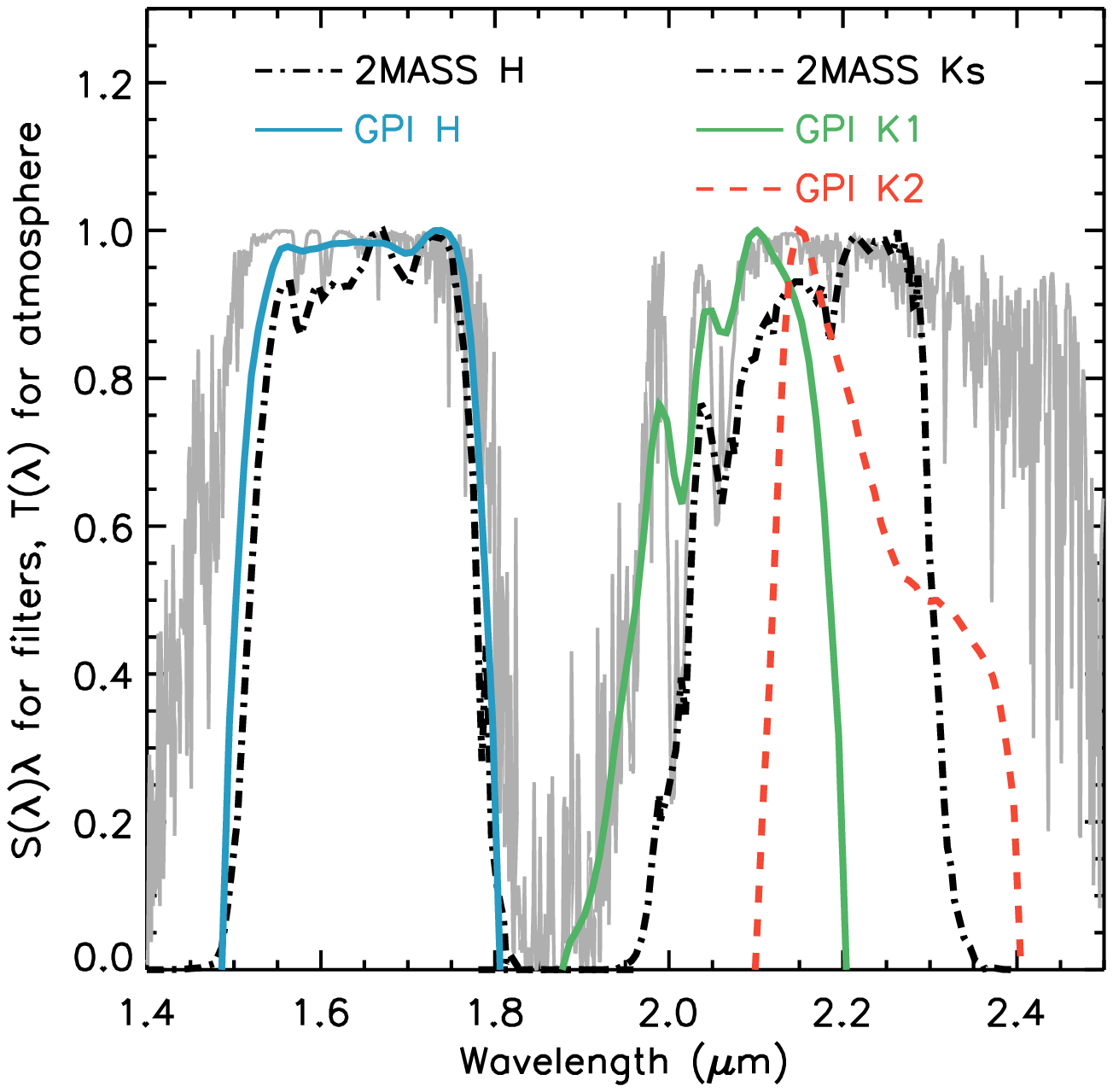}
\caption{The normalized relative spectral response $S(\lambda)\lambda$ of the {\it H}, {\it K$_1$}, and {\it K$_2$} GPI filters (blue solid, green solid, and red dashed curves), measured from coronagraphic observations of the white dwarf companion to HD 8049 \citep{Zurlo:2013kb}. $S(\lambda)\lambda$ for the 2MASS $H$ and $K_{\rm S}$ filters are also plotted (black dot-dashed curves; \citealp{Cohen:2003gg}). As in \citet{Tokunaga:2005ch}, $S(\lambda)$ includes the wavelength-dependent terrestrial atmospheric absorption, $T(\lambda)$, an example of which is plotted for Cerro Pach\'{o}n, assuming an airmass of 1.5 and a vertical water vapor column of 4.3~mm (thin gray curve; \citealp{Lord:1992to}).}
\label{fig:gpi_filters}
\end{figure}
As the photometric measurements of HD~95086~b used in this study were made using non-standard filters (GPI $H$, $K_1$, and NaCo $L^{\prime}$), the relative spectral response of each filter was obtained in order to extract synthetic photometry from model atmospheres, and a color transformation was calculated to convert between these filters, and the more standard MKO, 2MASS, and {\it WISE} photometric systems. For the GPI filters, the transmission curve and instrument throughput were measured using coronagraphic $H$, $K_1$, and $K_2$ observations of the white dwarf companion to HD~8049 \citep{Zurlo:2013kb}, and are plotted in Figure~\ref{fig:gpi_filters}. For the NaCo {\it L$^{\prime}$} filter, the transmission curve given in the instrument manual was multiplied by a {\tt SkyCalc}\footnote{\tt https://www.eso.org/observing/etc/skycalc} atmospheric transmission model for Cerro Paranal \citep{Noll:2012ch,Jones:2013ba}, computed at the average airmass of the NaCo observations of HD~95086~b, assuming a seasonal average for the precipitable water vapor. The instrument throughput of NaCo was assumed to be constant over the $L^{\prime}$ band-pass. For the 2MASS and {\it WISE}, the relative spectral response of the filters were obtained from \citet{Cohen:2003gg} and \citet{Wright:2010in}, respectively. The zero point of each filter was estimated by integrating the product of the flux-calibrated spectrum of Vega $F_{\lambda}(\lambda)$ \citep{Bohlin:2014bs}\footnote{\tt ftp://ftp.stsci.edu/cdbs/current\_calspec/\\alpha\_lyr\_stis\_008.fits} and the relative spectral response $S(\lambda)\lambda$ of the filter, divided by the integral of the relative spectral response, i.e.:
\begin{equation}
F_{\lambda} = \frac{\int F_{\lambda}(\lambda)S(\lambda)\lambda d\lambda}{\int S(\lambda)\lambda d\lambda}
\end{equation}
where $F_{\lambda}$ is the flux density of Vega for the filter, which is defined to have a magnitude of zero. The adopted zero points were: GPI $H$: $1.151\times 10^{-9}$~Wm$^{-2}$\micron$^{-1}$, $K_1$: $5.040\times 10^{-10}$~Wm$^{-2}$\micron$^{-1}$, $K_2$: $3.797\times 10^{-10}$~Wm$^{-2}$\micron$^{-1}$, and NaCo $L^{\prime}$: $5.127\times 10^{-11}$~Wm$^{-2}$\micron$^{-1}$. An uncertainty of 2\% on the zero point for each filter was assumed. The zero points for the 2MASS and {\it WISE} filters were consistent with the values reported in \citet{Cohen:2003gg} and \citet{Wright:2010in}.

\begin{figure}
\epsscale{0.5}
\plotone{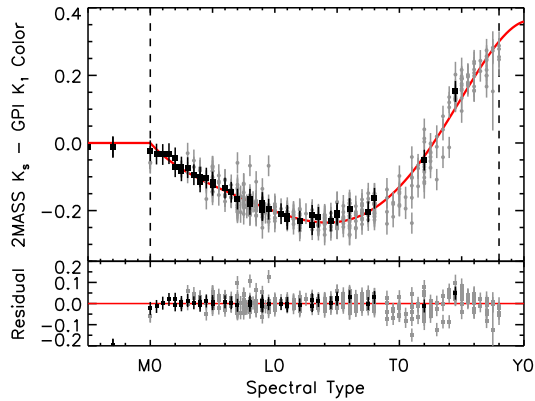}
\caption{Empirical color transformation between the 2MASS $K_{\rm S}$ and GPI $K_1$ filters, derived from objects within the IRTF Spectral Library (black squares), and the SpeX Prism Spectral Libraries (gray circles). The transformation is expressed as a fifth order polynomial over the M0--T8 spectral type range (red curve), the coefficients for which are given in Table~\ref{tab:filters}.}
\label{fig:filter_color}
\end{figure}
Empirical color transformations were computed between; the 2MASS \citep{Cohen:2003gg} and MKO \citep{Tokunaga:2002ex} photometric systems, and the GPI $H$, $K_1$, and $K_2$ filters; and between the MKO $L^{\prime}$ and {\it WISE} $W1$ filters. Empirical spectra of stars and brown dwarfs were obtained from the IRTF Spectral Library and the SpeX Prism Spectral Library. For each pair of filters for which a color transformation was required, the flux of the object in both filters was calculated (e.g., \citealp{Bessell:2012bq}), converted into a magnitude using the zero points described previously, and subtracted to compute the color. For each filter, Vega was assumed to have a magnitude of zero. To assess the uncertainty on the color transformation for a given object, the process was repeated $10^6$ times, each time drawing randomly from the uncertainty on the zero points of each filter, and the uncertainty on the flux of the object. This process was repeated for each object in both libraries, leading to an empirical color transformation as a function of spectral type, an example of which is shown in Figure~\ref{fig:filter_color}. For each object, the optical spectral type was preferentially used, although in some cases only an infrared estimate was available. A fifth order polynomial was fit to the color transformation for each filter pair over the M0--T8 spectral type range. As the color transformation was found to be negligible for stars of spectral type M0 and earlier, the first term of the polynomial was fixed to zero, such that the fit passed through zero for M0 spectral type. An example fit is shown in Figure~\ref{fig:filter_color}, and the polynomial coefficients for each filter pair are given in Table~\ref{tab:filters}.

\begin{deluxetable*}{cccccc}
\tablecaption{GPI filter relative spectral response}
\tablewidth{0pt}
\tablehead{
\multicolumn{2}{c}{$H$} & \multicolumn{2}{c}{$K_1$} & \multicolumn{2}{c}{$K_2$}\\
$\lambda$ (\micron) & $S(\lambda)\lambda$ & $\lambda$ (\micron) & $S(\lambda)\lambda$ & $\lambda$ (\micron) & $S(\lambda)\lambda$}
\startdata
1.4946&0.3327&1.8861&0.0376&2.1074&0.2099\\
1.5030&0.5142&1.8947&0.0539&2.1154&0.3846\\
1.5114&0.6807&1.9033&0.0745&2.1235&0.6079\\
1.5198&0.8029&1.9118&0.1082&2.1315&0.8342\\
1.5283&0.8679&1.9204&0.1550&2.1395&0.9576\\
1.5367&0.9134&1.9290&0.2197&2.1476&1.0000\\
1.5451&0.9513&1.9376&0.3010&2.1556&0.9943\\
1.5535&0.9749&1.9462&0.3698&2.1636&0.9611\\
1.5619&0.9783&1.9548&0.4358&2.1716&0.9211\\
1.5703&0.9748&1.9634&0.5021&2.1797&0.8799\\
1.5787&0.9716&1.9719&0.5988&2.1877&0.8501\\
1.5871&0.9743&1.9805&0.7119&2.1957&0.8153\\
1.5955&0.9754&1.9891&0.7655&2.2037&0.7885\\
1.6040&0.9788&1.9977&0.7436&2.2118&0.7550\\
1.6124&0.9818&2.0063&0.6658&2.2198&0.7103\\
1.6208&0.9823&2.0149&0.6268&2.2278&0.6745\\
1.6292&0.9825&2.0234&0.6942&2.2359&0.6473\\
1.6376&0.9843&2.0320&0.8139&2.2439&0.6027\\
1.6460&0.9844&2.0406&0.8905&2.2519&0.5747\\
1.6544&0.9823&2.0492&0.8909&2.2599&0.5509\\
1.6628&0.9831&2.0578&0.8625&2.2680&0.5300\\
1.6712&0.9828&2.0664&0.8599&2.2760&0.5258\\
1.6796&0.9798&2.0749&0.8955&2.2840&0.5178\\
1.6881&0.9738&2.0835&0.9502&2.2920&0.4992\\
1.6965&0.9688&2.0921&0.9905&2.3001&0.4973\\
1.7049&0.9701&2.1007&1.0000&2.3081&0.4991\\
1.7133&0.9796&2.1093&0.9896&2.3161&0.4899\\
1.7217&0.9927&2.1179&0.9688&2.3241&0.4854\\
1.7301&0.9997&2.1264&0.9526&2.3322&0.4728\\
1.7385&1.0000&2.1350&0.9347&2.3402&0.4586\\
1.7469&0.9951&2.1436&0.9100&2.3482&0.4435\\
1.7553&0.9794&2.1522&0.8776&2.3563&0.4308\\
1.7638&0.9373&2.1608&0.8242&2.3643&0.4158\\
1.7722&0.8557&2.1694&0.7436&2.3723&0.3959\\
1.7806&0.7153&2.1779&0.6285&2.3803&0.3586\\
1.7890&0.5268&2.1865&0.4750&2.3884&0.3013\\
1.7974&0.3366&2.1951&0.3185&2.3964&0.2272
\enddata
\label{tab:filtercurves}
\end{deluxetable*}
\begin{deluxetable*}{lccccccc}
\tablecaption{Empirical filter transformations}
\tablewidth{0pt}
\tablehead{
\colhead{Color} & \colhead{$c_0$}  & \colhead{$c_1$}  & \colhead{$c_2$}  & \colhead{$c_3$} & \colhead{$c_4$} & \colhead{$c_5$} & \colhead{rms}\\
&&$\times 10^{-2}$&$\times 10^{-3}$&$\times 10^{-4}$&$\times 10^{-6}$&$\times 10^{-7}$&}
\startdata

2MASS $H$ $-$ GPI $H$          &$0$&$-0.32073$          &$\phantom{-}0.22029$&$-0.13696$          &$\phantom{-}0.44056$&$-0.04090$&$0.003$\\
MKO $H$ $-$ GPI $H$            &$0$&$\phantom{-}0.52976$&$-0.45040$          &$\phantom{-}0.35256$&$-1.58498$          &$\phantom{-}0.26114$&$0.004$\\
2MASS $K_{\rm S}$ $-$ GPI $K_1$&$0$&$-4.06008$          &$\phantom{-}5.24441$&$-5.47733$          &$\phantom{-}26.8625$&$-4.16143$&$0.031$\\
2MASS $K_{\rm S}$ $-$ GPI $K_2$&$0$&$\phantom{-}2.55246$&$-3.00911$          &$\phantom{-}2.23060$&$-7.68838$          &$\phantom{-}0.55557$&$0.027$\\
MKO $K$ $-$ GPI $K_1$          &$0$&$-5.81313$          &$\phantom{-}8.81710$&$-8.88431$          &$\phantom{-}41.3378$&$-6.23493$&$0.040$\\
MKO $K$ $-$ GPI $K_2$          &$0$&$\phantom{-}0.74663$&$\phantom{-}0.70976$&$-1.31673$          &$\phantom{-}7.34113$&$-1.59507$&$0.019$\\
MKO $K^{\prime}$ $-$ GPI $K_1$ &$0$&$-2.53922$          &$\phantom{-}3.55111$&$-3.81046$          &$\phantom{-}18.9157$&$-2.91471$&$0.023$\\
MKO $K^{\prime}$ $-$ GPI $K_2$ &$0$&$\phantom{-}3.98467$&$-4.42024$          &$\phantom{-}3.59249$&$-14.2935$          &$\phantom{-}1.59708$&$0.033$\\
MKO $K_{\rm S}$ $-$ GPI $K_1$  &$0$&$-3.59593$          &$\phantom{-}4.61246$&$-4.85141$          &$\phantom{-}23.9600$&$-3.73692$&$0.027$\\
MKO $K_{\rm S}$ $-$ GPI $K_2$  &$0$&$\phantom{-}2.90088$&$-3.28302$          &$\phantom{-}2.47902$&$-8.95925$          &$\phantom{-}0.73357$&$0.030$\\
\hline
MKO $L^{\prime}$ $-$ {\it WISE} $W1$ &$0$&$\phantom{-}0.99889$&$-15.2677$&$\phantom{-}15.7383$&$-65.6107$&$\phantom{-}9.14685$&$0.113$
\enddata
\tablecomments{Polynomial coefficients are defined as $y=\sum_i c_i x^i$, where $y$ is the color transform, $x$ is the spectral type. Spectral types are defined such that M0 = 0, L0 = 10, T0 = 20.}
\label{tab:filters}
\end{deluxetable*}
\begin{deluxetable*}{lccccccc}
\tablecaption{Properties of young low-mass companions and isolated young and/or dusty brown dwarfs used as comparison objects}
\tablewidth{0pt}
\tablehead{
\colhead{Name} & \colhead{Age} & \colhead{Spectral Type} & \colhead{$T_{\rm eff}$} & \colhead{$\log g$} & \colhead{$\log L/L_\odot$} & \colhead{Mass} & \colhead{References} \\
\colhead{} & \colhead{(Myr)} & \colhead{} & \colhead{(K)} & \colhead{(dex)} & \colhead{(dex)} & \colhead{($M_{\rm Jup}$)} & \colhead{}}
\startdata
HIP\,78530\,B & $5$ & M$7_\beta\pm0.5$ & $2700 \pm100$ & $4.5\pm1.0 $ & $-2.53\pm0.09 $ & $22\pm 1$ & {1, 2} \\
GSC\,06214-0010\,B & $5$ & M$9_\gamma\pm0.5$ & $2300 \pm100$ & $3\pm0.5 $ & $-3.01\pm0.09 $ & $15\pm 1$ & {2, 3, 4} \\
1\,RXS\,J160929.1-210524\,b & $5$ & L$4_\gamma\pm1$ & $1700\pm100$ & $3.5\pm0.5 $ & $-3.5\pm0.2 $ & $8\pm 1$ & {2, 5} \\
2MASS\,J120734-393253\,b & $8$ & $>\mathrm{L}5_\gamma$ & $1600\pm100$ & $4.0\pm0.5$ & $-4.72\pm0.14 $ & $8\pm 2$ & {6, 7, 8, 9} \\
2MASS~J11193254-1137466 & $10\pm3$ & L7 & -- & -- & -- & 5--6 & {10} \\
PSO~J318.5338-22.8603 & $12_{-4}^{+8}$ & L$7_{\gamma}\pm1$ & $1160^{+30}_{-40}$ & $3.86_{-0.08}^{+0.10}$ & $-4.42 \pm 0.06$ & $6.5^{+1.3}_{-1.0}$ & {11}\\
HIP\,106906\,b & $13\pm2$ & L$2.5\pm1$ & $1800\pm100$ & -- & $-3.64\pm0.08 $ & $11\pm 2$ & {12} \\
HR\,8799\,b & $30$ & L--T & $750-1100$ & $4.0\pm0.5$ & $-5.1\pm0.1 $ & $\sim5$ & {13,14,15} \\
HR\,8799\,c & $30$ & L--T & $1100\pm100$ & $\sim4.0 $ & $-4.7\pm0.1 $ & $\sim7$ & {14,15} \\
HR\,8799\,d & $30$ & L--T & $1100\pm100 $ & $\sim4.0 $ & $-4.7\pm0.1 $ & $\sim7$ & {14,15} \\
HR\,8799\,e & $30$ & L--T & $1100\pm100$ & $\sim4.0$ & $-4.7\pm0.2 $ & $\sim7$ & {16}\\
2MASS~J21481628+4003593 & -- & L6\tablenotemark{\it a} & 1500 & 3.0 & $-4.07\pm0.12$ & -- & {17, 18}\\
2MASS~J22443167+2043433 & -- & L6.5\tablenotemark{\it a} & -- & -- & -- & -- & {19}
\enddata
\tablenotetext{a}{Optical spectral type}
\label{tab:obj_comp}
\tablerefs{(1) \citealp{Lafreniere:2011dh}; (2) \citealp{Lachapelle:2015cx}; (3) \citealp{Ireland:2011id}; (4) \citealp{Bowler:2011gw}; (5) \citealp{Lafreniere:2010cp};  (6) \citealp{Chauvin:2004cy}; (7) \citealp{Patience:2010hf};  (8) \citealp{Faherty:2013bc}; (9) \citealp{Mohanty:2007er}; (10) \citealp{Kellogg:2015cf}; (11) \citealp{Liu:2013gy}; (12) \citealp{Bailey:2014et}; (13) \citealp{Marois:2008ei}; (14) \citealp{Barman:2011fe}; (15) \citealp{Marley:2012fo}; (16) \citealp{Marois:2010gp}; (17) \citealp{Looper:2008hs}; (18) \citealp{Witte:2011kn}; (19) \citealp{Dahn:2002fu}.}
\end{deluxetable*}

\clearpage


\begin{thebibliography}{}
\expandafter\ifx\csname natexlab\endcsname\relax\def\natexlab#1{#1}\fi

\bibitem[{Allard {et~al.}(2001)Allard, Hauschildt, Alexander, Tamanai, \&
  Schweitzer}]{Allard:2001fh}
Allard, F., Hauschildt, P.~H., Alexander, D.~R., Tamanai, A., \& Schweitzer, A.
  2001, ApJ, 556, 357

\bibitem[{Allard {et~al.}(2012)Allard, Homeier, \& Freytag}]{Allard:2012fp}
Allard, F., Homeier, D., \& Freytag, B. 2012, Philosophical Transactions of the
  Royal Society A: Mathematical, Physical and Engineering Sciences, 370, 2765

\bibitem[{Allende~Prieto \& Lambert(1999)}]{AllendePrieto:1999td}
Allende~Prieto, C., \& Lambert, D.~L. 1999, A{\&}A, 352, 555

\bibitem[{Allers \& Liu(2013)}]{Allers:2013hk}
Allers, K.~N., \& Liu, M.~C. 2013, ApJ, 772, 79

\bibitem[{Asplund {et~al.}(2009)Asplund, Grevesse, Sauval, \&
  Scott}]{Asplund:2009eu}
Asplund, M., Grevesse, N., Sauval, A.~J., \& Scott, P. 2009, ARA{\&}A, 47, 481

\bibitem[{Bailey {et~al.}(2013)Bailey, Hinz, Currie, Su, Esposito, Hill,
  Hoffmann, Jones, Kim, Leisenring, Meyer, Murray-Clay, Nelson, Pinna, Puglisi,
  Rieke, Rodigas, Skemer, Skrutskie, Vaitheeswaran, \& Wilson}]{Bailey:2013gl}
Bailey, V., Hinz, P.~M., Currie, T., {et~al.} 2013, ApJ, 767, 31

\bibitem[{Bailey {et~al.}(2014)Bailey, Meshkat, Reiter, Morzinski, Males, Su,
  Hinz, Kenworthy, Stark, Mamajek, Briguglio, Close, Follette, Puglisi,
  Rodigas, Weinberger, \& Xompero}]{Bailey:2014et}
Bailey, V., Meshkat, T., Reiter, M., {et~al.} 2014, ApJ, 780, L4

\bibitem[{Baraffe {et~al.}(2003)Baraffe, Chabrier, Barman, Allard, \&
  Hauschildt}]{Baraffe:2003bj}
Baraffe, I., Chabrier, G., Barman, T.~S., Allard, F., \& Hauschildt, P.~H.
  2003, A{\&}A, 402, 701

\bibitem[{Barber {et~al.}(2006)Barber, Tennyson, Harris, \&
  Tolchenov}]{Barber:2006dm}
Barber, R.~J., Tennyson, J., Harris, G.~J., \& Tolchenov, R.~N. 2006, MNRAS,
  368, 1087

\bibitem[{Barman {et~al.}(2011{\natexlab{a}})Barman, Macintosh, Konopacky, \&
  Marois}]{Barman:2011fe}
Barman, T.~S., Macintosh, B., Konopacky, Q.~M., \& Marois, C.
  2011{\natexlab{a}}, ApJ, 733, 65

\bibitem[{Barman {et~al.}(2011{\natexlab{b}})Barman, Macintosh, Konopacky, \&
  Marois}]{Barman:2011dq}
---. 2011{\natexlab{b}}, ApJL, 735, L39

\bibitem[{Baudino {et~al.}(2015)Baudino, B{\'e}zard, Boccaletti, Bonnefoy,
  Lagrange, \& Galicher}]{Baudino:2015kh}
Baudino, J.~L., B{\'e}zard, B., Boccaletti, A., {et~al.} 2015, A{\&}A, 582, A83

\bibitem[{Bean {et~al.}(2013)Bean, Desert, Seifahrt, Madhusudhan, Chilingarian,
  Homeier, \& Szentgyorgyi}]{Bean:2013dg}
Bean, J.~L., Desert, J.-M., Seifahrt, A., {et~al.} 2013, ApJ, 771, 108

\bibitem[{Bessell \& Murphy(2012)}]{Bessell:2012bq}
Bessell, M., \& Murphy, S. 2012, PASP, 124, 140

\bibitem[{Bohlin(2014)}]{Bohlin:2014bs}
Bohlin, R.~C. 2014, AJ, 147, 127

\bibitem[{Bonneau {et~al.}(2006)Bonneau, Clausse, Delfosse, Mourard, Cetre,
  Chelli, Cruzal{\`e}bes, Duvert, \& Zins}]{Bonneau:2006bl}
Bonneau, D., Clausse, J.~M., Delfosse, X., {et~al.} 2006, A{\&}A, 456, 789

\bibitem[{Bonnefoy {et~al.}(2014{\natexlab{a}})Bonnefoy, Chauvin, Lagrange,
  Rojo, Allard, Pinte, Dumas, \& Homeier}]{Bonnefoy:2014dh}
Bonnefoy, M., Chauvin, G., Lagrange, A.-M., {et~al.} 2014{\natexlab{a}},
  A{\&}A, 562, 127

\bibitem[{Bonnefoy {et~al.}(2014{\natexlab{b}})Bonnefoy, Marleau, Galicher,
  Beust, Lagrange, Baudino, Chauvin, Borgniet, Meunier, Rameau, Boccaletti,
  Cumming, Helling, Homeier, Allard, \& Delorme}]{Bonnefoy:2014bx}
Bonnefoy, M., Marleau, G.~D., Galicher, R., {et~al.} 2014{\natexlab{b}},
  A{\&}A, 567, L9

\bibitem[{Bowler {et~al.}(2010)Bowler, Liu, Dupuy, \& Cushing}]{Bowler:2010ft}
Bowler, B.~P., Liu, M.~C., Dupuy, T.~J., \& Cushing, M.~C. 2010, ApJ, 723, 850

\bibitem[{{Bowler} {et~al.}(2014){Bowler}, {Liu}, {Kraus}, \&
  {Mann}}]{Bowler:2014}
{Bowler}, B.~P., {Liu}, M.~C., {Kraus}, A.~L., \& {Mann}, A.~W. 2014, \apj,
  784, 65

\bibitem[{Bowler {et~al.}(2011)Bowler, Liu, Kraus, Mann, \&
  Ireland}]{Bowler:2011gw}
Bowler, B.~P., Liu, M.~C., Kraus, A.~L., Mann, A.~W., \& Ireland, M.~J. 2011,
  ApJ, 743, 148

\bibitem[{Burrows {et~al.}(2006)Burrows, Sudarsky, \& Hubeny}]{Burrows:2006ia}
Burrows, A., Sudarsky, D., \& Hubeny, I. 2006, ApJ, 640, 1063

\bibitem[{Burrows {et~al.}(1997)Burrows, Marley, Hubbard, Lunine, Guillot,
  Saumon, Freedman, Sudarsky, \& Sharp}]{Burrows:1997ua}
Burrows, A., Marley, M., Hubbard, W.~B., {et~al.} 1997, ApJ, 491, 856

\bibitem[{Caffau {et~al.}(2011)Caffau, Ludwig, Steffen, Freytag, \&
  Bonifacio}]{Caffau:2011ik}
Caffau, E., Ludwig, H.~G., Steffen, M., Freytag, B., \& Bonifacio, P. 2011,
  Solar Physics, 268, 255

\bibitem[{Chabrier {et~al.}(2000)Chabrier, Baraffe, Allard, \&
  Hauschildt}]{Chabrier:2000hq}
Chabrier, G., Baraffe, I., Allard, F., \& Hauschildt, P. 2000, ApJ, 542, 464

\bibitem[{Chauvin {et~al.}(2004)Chauvin, Lagrange, Dumas, Zuckerman, Mouillet,
  Song, Beuzit, \& Lowrance}]{Chauvin:2004cy}
Chauvin, G., Lagrange, A.-M., Dumas, C., {et~al.} 2004, A{\&}A, 425, L29

\bibitem[{Chen {et~al.}(2012)Chen, Pecaut, Mamajek, Su, \&
  Bitner}]{Chen:2012ki}
Chen, C.~H., Pecaut, M., Mamajek, E.~E., Su, K. Y.~L., \& Bitner, M. 2012, ApJ,
  756, 133

\bibitem[{Cohen {et~al.}(2003)Cohen, Wheaton, \& Megeath}]{Cohen:2003gg}
Cohen, M., Wheaton, W.~A., \& Megeath, S.~T. 2003, AJ, 126, 1090

\bibitem[{Cruz {et~al.}(2009)Cruz, Kirkpatrick, \& Burgasser}]{Cruz:2009gs}
Cruz, K.~L., Kirkpatrick, J.~D., \& Burgasser, A.~J. 2009, AJ, 137, 3345

\bibitem[{Currie {et~al.}(2013)Currie, Burrows, Madhusudhan, Fukagawa, Girard,
  Dawson, Murray-Clay, Kenyon, Kuchner, Matsumura, Jayawardhana, Chambers, \&
  Bromley}]{Currie:2013cv}
Currie, T., Burrows, A., Madhusudhan, N., {et~al.} 2013, ApJ, 776, 15

\bibitem[{Cushing {et~al.}(2005)Cushing, Rayner, \& Vacca}]{Cushing:2005ed}
Cushing, M.~C., Rayner, J.~T., \& Vacca, W.~D. 2005, ApJ, 623, 1115

\bibitem[{Cutri(2014)}]{Cutri:2014wx}
Cutri, R.~M. 2014, VizieR Online Data Catalog, 2328, 0

\bibitem[{Dahn {et~al.}(2002)Dahn, Harris, Vrba, Guetter, Canzian, Henden,
  Levine, Luginbuhl, Monet, Monet, Pier, Stone, Walker, Burgasser, Gizis,
  Kirkpatrick, Liebert, \& Reid}]{Dahn:2002fu}
Dahn, C.~C., Harris, H.~C., Vrba, F.~J., {et~al.} 2002, AJ, 124, 1170

\bibitem[{de~Zeeuw {et~al.}(1999)de~Zeeuw, Hoogerwerf, de~Bruijne, Brown, \&
  Blaauw}]{deZeeuw:1999fe}
de~Zeeuw, P.~T., Hoogerwerf, R., de~Bruijne, J. H.~J., Brown, A. G.~A., \&
  Blaauw, A. 1999, AJ, 117, 354

\bibitem[{Delorme {et~al.}(2013)Delorme, Gagn{\'e}, Girard, Lagrange, Chauvin,
  Naud, Lafreni{\`e}re, Doyon, Riedel, Bonnefoy, \& Malo}]{Delorme:2013bo}
Delorme, P., Gagn{\'e}, J., Girard, J.~H., {et~al.} 2013, A{\&}A, 553, L5

\bibitem[{Deming {et~al.}(2013)Deming, Wilkins, McCullough, Burrows, Fortney,
  Agol, Dobbs-Dixon, Madhusudhan, Crouzet, Desert, Gilliland, Haynes, Knutson,
  Line, Magic, Mandell, Ranjan, Charbonneau, Clampin, Seager, \&
  Showman}]{Deming:2013ge}
Deming, D., Wilkins, A., McCullough, P., {et~al.} 2013, ApJ, 774, 95

\bibitem[{Dotter {et~al.}(2008)Dotter, Chaboyer, Jevremovi{\'c}, Kostov, Baron,
  \& Ferguson}]{Dotter:2008ga}
Dotter, A., Chaboyer, B., Jevremovi{\'c}, D., {et~al.} 2008, ApJS, 178, 89

\bibitem[{Dupuy \& Liu(2012)}]{Dupuy:2012bp}
Dupuy, T.~J., \& Liu, M.~C. 2012, ApJS, 201, 19

\bibitem[{Esposito {et~al.}(2013)Esposito, Mesa, Skemer, Arcidiacono, Claudi,
  Desidera, Gratton, Mannucci, Marzari, Masciadri, Close, Hinz, Kulesa,
  McCarthy, Males, Agapito, Argomedo, Boutsia, Briguglio, Brusa, Busoni,
  Cresci, Fini, Fontana, Guerra, Hill, Miller, Paris, Pinna, Puglisi,
  Quiros-Pacheco, Riccardi, Stefanini, Testa, Xompero, \&
  Woodward}]{Esposito:2013hs}
Esposito, S., Mesa, D., Skemer, A., {et~al.} 2013, A{\&}A, 549, 52

\bibitem[{Faherty {et~al.}(2013)Faherty, Rice, Cruz, Mamajek, \&
  N{\'u}{\~n}ez}]{Faherty:2013bc}
Faherty, J.~K., Rice, E.~L., Cruz, K.~L., Mamajek, E.~E., \& N{\'u}{\~n}ez, A.
  2013, AJ, 145, 2

\bibitem[{Faherty {et~al.}(2012)Faherty, Burgasser, Walter, Van~der Bliek,
  Shara, Cruz, West, Vrba, \& Anglada-Escud{\'e}}]{Faherty:2012cy}
Faherty, J.~K., Burgasser, A.~J., Walter, F.~M., {et~al.} 2012, ApJ, 752, 56

\bibitem[{Filippazzo {et~al.}(2015)Filippazzo, Rice, Faherty, Cruz, Van~Gordon,
  \& Looper}]{Filippazzo:2015dv}
Filippazzo, J.~C., Rice, E.~L., Faherty, J., {et~al.} 2015, ApJ, 810, 158

\bibitem[{Foreman-Mackey {et~al.}(2013)Foreman-Mackey, Hogg, Lang, \&
  Goodman}]{ForemanMackey:2013io}
Foreman-Mackey, D., Hogg, D.~W., Lang, D., \& Goodman, J. 2013, PASP, 125, 306

\bibitem[{Gagn{\'e} {et~al.}(2015)Gagn{\'e}, Burgasser, Faherty,
  Lafreni{\`e}re, Doyon, Filippazzo, Bowsher, \& Nicholls}]{Gagne:2015kf}
Gagn{\'e}, J., Burgasser, A.~J., Faherty, J.~K., {et~al.} 2015, ApJL, 808, L20

\bibitem[{Galicher {et~al.}(2014)Galicher, Rameau, Bonnefoy, Baudino, Currie,
  Boccaletti, Chauvin, Lagrange, \& Marois}]{Galicher:2014er}
Galicher, R., Rameau, J., Bonnefoy, M., {et~al.} 2014, 565, L4

\bibitem[{Gauza {et~al.}(2015)Gauza, Bejar, P{\'e}rez-Garrido, Rosa
  Zapatero~Osorio, Lodieu, Rebolo, Pall{\'e}, \& Nowak}]{Gauza:2015fw}
Gauza, B., Bejar, V. J.~S., P{\'e}rez-Garrido, A., {et~al.} 2015, ApJ, 804, 96

\bibitem[{Golimowski {et~al.}(2004)Golimowski, Leggett, Marley, Fan, Geballe,
  Knapp, Vrba, Henden, Luginbuhl, Guetter, Munn, Canzian, Zheng, Tsvetanov,
  Chiu, Glazebrook, Hoversten, Schneider, \& Brinkmann}]{Golimowski:2004en}
Golimowski, D.~A., Leggett, S.~K., Marley, M.~S., {et~al.} 2004, AJ, 127, 3516

\bibitem[{Greco \& Brandt(2016)}]{Greco:2016ww}
Greco, J.~P., \& Brandt, T.~D. 2016, eprint arXiv:1602.00691, 1602.00691

\bibitem[{Grevesse {et~al.}(1993)Grevesse, Noels, \& Sauval}]{Grevesse:1993uj}
Grevesse, N., Noels, A., \& Sauval, A.~J. 1993, A{\&}A, 271, 587

\bibitem[{Hauschildt(1992)}]{Hauschildt:1992ff}
Hauschildt, P.~H. 1992, Journal of Quantitative Spectroscopy and Radiative
  Transfer, 47, 433

\bibitem[{Helling {et~al.}(2008)Helling, Dehn, Woitke, \&
  Hauschildt}]{Helling:2008ht}
Helling, C., Dehn, M., Woitke, P., \& Hauschildt, P.~H. 2008, ApJL, 675, L105

\bibitem[{Helling \& Woitke(2006)}]{Helling:2006gp}
Helling, C., \& Woitke, P. 2006, A{\&}A, 455, 325

\bibitem[{Hinz {et~al.}(2010)Hinz, Rodigas, Kenworthy, Sivanandam, Heinze,
  Mamajek, \& Meyer}]{Hinz:2010fy}
Hinz, P.~M., Rodigas, T.~J., Kenworthy, M.~A., {et~al.} 2010, ApJ, 716, 417

\bibitem[{H{\o}g {et~al.}(2000)H{\o}g, Fabricius, Makarov, Urban, Corbin,
  Wycoff, Bastian, Schwekendiek, \& Wicenec}]{Hog:2000wk}
H{\o}g, E., Fabricius, C., Makarov, V.~V., {et~al.} 2000, 355, L27

\bibitem[{Hubeny(1988)}]{Hubeny:1988ci}
Hubeny, I. 1988, Computer Physics Communications, 52, 103

\bibitem[{Hubeny \& Burrows(2007)}]{Hubeny:2007hm}
Hubeny, I., \& Burrows, A. 2007, ApJ, 669, 1248

\bibitem[{Hubeny {et~al.}(2003)Hubeny, Burrows, \& Sudarsky}]{Hubeny:2003eb}
Hubeny, I., Burrows, A., \& Sudarsky, D. 2003, ApJ, 594, 1011

\bibitem[{Hubeny \& Lanz(1995)}]{Hubeny:1995fw}
Hubeny, I., \& Lanz, T. 1995, ApJ, 439, 875

\bibitem[{Ingraham {et~al.}(2014)Ingraham, Marley, Saumon, Marois, Macintosh,
  Barman, Bauman, Burrows, Chilcote, De~Rosa, Dillon, Doyon, Dunn, Erikson,
  Fitzgerald, Gavel, Goodsell, Graham, Hartung, Hibon, Kalas, Konopacky,
  Larkin, Maire, Marchis, McBride, Millar-Blanchaer, Morzinski, Norton,
  Oppenheimer, Palmer, Patience, Perrin, Poyneer, Pueyo, Rantakyro, Sadakuni,
  Saddlemyer, Savransky, Soummer, Sivaramakrishnan, Song, Thomas, Wallace,
  Wiktorowicz, \& Wolff}]{Ingraham:2014gx}
Ingraham, P., Marley, M.~S., Saumon, D., {et~al.} 2014, ApJ, 794, L15

\bibitem[{Ireland {et~al.}(2011)Ireland, Kraus, Martinache, Law, \&
  Hillenbrand}]{Ireland:2011id}
Ireland, M.~J., Kraus, A., Martinache, F., Law, N., \& Hillenbrand, L.~A. 2011,
  ApJ, 726, 113

\bibitem[{Janson {et~al.}(2010)Janson, Bergfors, Goto, Brandner, \&
  Lafreni{\`e}re}]{Janson:2010db}
Janson, M., Bergfors, C., Goto, M., Brandner, W., \& Lafreni{\`e}re, D. 2010,
  ApJL, 710, L35

\bibitem[{Jones {et~al.}(2013)Jones, Noll, Kausch, Szyszka, \&
  Kimeswenger}]{Jones:2013ba}
Jones, A., Noll, S., Kausch, W., Szyszka, C., \& Kimeswenger, S. 2013, A{\&}A,
  560, 91

\bibitem[{Kellogg {et~al.}(2015)Kellogg, Metchev, Gei{\ss}ler, Hicks,
  Kirkpatrick, \& Kurtev}]{Kellogg:2015cf}
Kellogg, K., Metchev, S., Gei{\ss}ler, K., {et~al.} 2015, AJ, 150, 182

\bibitem[{Kirkpatrick(2005)}]{Kirkpatrick:2005cv}
Kirkpatrick, J.~D. 2005, ARA{\&}A, 43, 195

\bibitem[{Knutson {et~al.}(2008)Knutson, Charbonneau, Allen, Burrows, \&
  Megeath}]{Knutson:2008gl}
Knutson, H.~A., Charbonneau, D., Allen, L.~E., Burrows, A., \& Megeath, S.~T.
  2008, ApJ, 673, 526

\bibitem[{Konopacky {et~al.}(2013)Konopacky, Barman, Macintosh, \&
  Marois}]{Konopacky:2013jv}
Konopacky, Q.~M., Barman, T.~S., Macintosh, B.~A., \& Marois, C. 2013, Science,
  339, 1398

\bibitem[{Konopacky {et~al.}(2014)Konopacky, Thomas, Macintosh, Dillon,
  Sadakuni, Maire, Fitzgerald, Hinkley, Kalas, Esposito, Marois, Ingraham,
  Marchis, Perrin, Graham, Wang, De~Rosa, Morzinski, Pueyo, Chilcote, Larkin,
  Fabrycky, Goodsell, Oppenheimer, Patience, Saddlemyer, \&
  Sivaramakrishnan}]{Konopacky:2014hf}
Konopacky, Q.~M., Thomas, S.~J., Macintosh, B.~A., {et~al.} 2014, Proc. SPIE,
  9147, 84

\bibitem[{Kraus {et~al.}(2014)Kraus, Ireland, Cieza, Hinkley, Dupuy, Bowler, \&
  Liu}]{Kraus:2014hk}
Kraus, A.~L., Ireland, M.~J., Cieza, L.~A., {et~al.} 2014, ApJ, 781, 20

\bibitem[{Lachapelle {et~al.}(2015)Lachapelle, Lafreni{\`e}re, Gagn{\'e},
  Jayawardhana, Janson, Helling, \& Witte}]{Lachapelle:2015cx}
Lachapelle, F.-R., Lafreni{\`e}re, D., Gagn{\'e}, J., {et~al.} 2015, ApJ, 802,
  61

\bibitem[{Lafreni{\`e}re {et~al.}(2011)Lafreni{\`e}re, Jayawardhana, Janson,
  Helling, Witte, \& Hauschildt}]{Lafreniere:2011dh}
Lafreni{\`e}re, D., Jayawardhana, R., Janson, M., {et~al.} 2011, ApJ, 730, 42

\bibitem[{Lafreni{\`e}re {et~al.}(2008)Lafreni{\`e}re, Jayawardhana, \& van
  Kerkwijk}]{Lafreniere:2008jt}
Lafreni{\`e}re, D., Jayawardhana, R., \& van Kerkwijk, M.~H. 2008, ApJL, 689,
  L153

\bibitem[{Lafreni{\`e}re {et~al.}(2010)Lafreni{\`e}re, Jayawardhana, \& van
  Kerkwijk}]{Lafreniere:2010cp}
---. 2010, ApJ, 719, 497

\bibitem[{Lafreni{\`e}re {et~al.}(2007)Lafreni{\`e}re, Marois, Doyon, Nadeau,
  \& Artigau}]{Lafreniere:2007bg}
Lafreni{\`e}re, D., Marois, C., Doyon, R., Nadeau, D., \& Artigau, {\'E}. 2007,
  ApJ, 660, 770

\bibitem[{Lagrange {et~al.}(2009)Lagrange, Gratadour, Chauvin, Fusco,
  Ehrenreich, Mouillet, Rousset, Rouan, Allard, Gendron, Charton, Mugnier,
  Rabou, Montri, \& Lacombe}]{Lagrange:2009hq}
Lagrange, A.-M., Gratadour, D., Chauvin, G., {et~al.} 2009, 493, L21

\bibitem[{Lagrange {et~al.}(2010)Lagrange, Bonnefoy, Chauvin, Apai, Ehrenreich,
  Boccaletti, Gratadour, Rouan, Mouillet, Lacour, \& Kasper}]{Lagrange:2010fs}
Lagrange, A.-M., Bonnefoy, M., Chauvin, G., {et~al.} 2010, Science, 329, 57

\bibitem[{Leggett {et~al.}(2007)Leggett, Saumon, Marley, Geballe, Golimowski,
  Stephens, \& Fan}]{Leggett:2007if}
Leggett, S.~K., Saumon, D., Marley, M.~S., {et~al.} 2007, ApJ, 655, 1079

\bibitem[{Leggett {et~al.}(2003)Leggett, Hawarden, Currie, Adamson, Carroll,
  Kerr, Kuhn, Seigar, Varricatt, \& Wold}]{Leggett:2003gt}
Leggett, S.~K., Hawarden, T.~G., Currie, M.~J., {et~al.} 2003, MNRAS, 345, 144

\bibitem[{Leggett {et~al.}(2010)Leggett, Burningham, Saumon, Marley, Warren,
  Smart, Jones, Lucas, Pinfield, \& Tamura}]{Leggett:2010cl}
Leggett, S.~K., Burningham, B., Saumon, D., {et~al.} 2010, ApJ, 710, 1627

\bibitem[{Liu {et~al.}(2010)Liu, Dupuy, \& Leggett}]{Liu:2010cw}
Liu, M.~C., Dupuy, T.~J., \& Leggett, S.~K. 2010, ApJ, 722, 311

\bibitem[{Liu {et~al.}(2013)Liu, Magnier, Deacon, Allers, Dupuy, Kotson, Aller,
  Burgett, Chambers, Draper, Hodapp, Jedicke, Kaiser, Kudritzki, Metcalfe,
  Morgan, Price, Tonry, \& Wainscoat}]{Liu:2013gy}
Liu, M.~C., Magnier, E.~A., Deacon, N.~R., {et~al.} 2013, ApJL, 777, L20

\bibitem[{Looper {et~al.}(2008)Looper, Kirkpatrick, Cutri, Barman, Burgasser,
  Cushing, Roellig, McGovern, McLean, Rice, Swift, \& Schurr}]{Looper:2008hs}
Looper, D.~L., Kirkpatrick, J.~D., Cutri, R.~M., {et~al.} 2008, ApJ, 686, 528

\bibitem[{Lord(1992)}]{Lord:1992to}
Lord, S.~D. 1992, NASA Technical Memorandum 103957, -1

\bibitem[{Macintosh {et~al.}(2014)Macintosh, Graham, Ingraham, Konopacky,
  Marois, Perrin, Poyneer, Bauman, Barman, Burrows, Cardwell, Chilcote,
  De~Rosa, Dillon, Doyon, Dunn, Erikson, Fitzgerald, Gavel, Goodsell, Hartung,
  Hibon, Kalas, Larkin, Maire, Marchis, Marley, McBride, Millar-Blanchaer,
  Morzinski, Norton, Oppenheimer, Palmer, Patience, Pueyo, Rantakyro, Sadakuni,
  Saddlemyer, Savransky, Serio, Soummer, Sivaramakrishnan, Song, Thomas,
  Wallace, Wiktorowicz, \& Wolff}]{Macintosh:2014js}
Macintosh, B., Graham, J.~R., Ingraham, P., {et~al.} 2014, PNAS, 111, 12661

\bibitem[{Macintosh {et~al.}(2015)Macintosh, Graham, Barman, De~Rosa,
  Konopacky, Marley, Marois, Nielsen, Pueyo, Rajan, Rameau, Saumon, Wang,
  Patience, Ammons, Arriaga, Artigau, Beckwith, Brewster, Bruzzone, Bulger,
  Burningham, Burrows, Chen, Chiang, Chilcote, Dawson, Dong, Doyon, Draper,
  Duch{\^e}ne, Esposito, Fabrycky, Fitzgerald, Follette, Fortney, Gerard,
  Goodsell, Greenbaum, Hibon, Hinkley, Cotten, Hung, Ingraham, Johnson-Groh,
  Kalas, Lafreni{\`e}re, Larkin, Lee, Line, Long, Maire, Marchis, Matthews,
  Max, Metchev, Millar-Blanchaer, Mittal, Morley, Morzinski, Murray-Clay,
  Oppenheimer, Palmer, Patel, Perrin, Poyneer, Rafikov, Rantakyr{\"o}, Rice,
  Rojo, Rudy, Ruffio, Ruiz, Sadakuni, Saddlemyer, Salama, Savransky, Schneider,
  Sivaramakrishnan, Song, Soummer, Thomas, Vasisht, Wallace, Ward-Duong,
  Wiktorowicz, Wolff, \& Zuckerman}]{Macintosh:2015ew}
Macintosh, B., Graham, J.~R., Barman, T., {et~al.} 2015, Science, 350, 64

\bibitem[{Madhusudhan {et~al.}(2011)Madhusudhan, Burrows, \&
  Currie}]{Madhusudhan:2011ex}
Madhusudhan, N., Burrows, A., \& Currie, T. 2011, ApJ, 737, 34

\bibitem[{Madsen {et~al.}(2002)Madsen, Dravins, \& Lindegren}]{Madsen:2002ha}
Madsen, S., Dravins, D., \& Lindegren, L. 2002, A{\&}A, 381, 446

\bibitem[{Maire {et~al.}(2014)Maire, Ingraham, De~Rosa, Perrin, Rajan,
  Savransky, Wang, Ruffio, Wolff, Chilcote, Doyon, Graham, Greenbaum,
  Konopacky, Larkin, Macintosh, Marois, Millar-Blanchaer, Patience, Pueyo,
  Sivaramakrishnan, Thomas, \& Weiss}]{Maire:2014gs}
Maire, J., Ingraham, P.~J., De~Rosa, R.~J., {et~al.} 2014, Proc. SPIE, 9147, 85

\bibitem[{Mamajek {et~al.}(2002)Mamajek, Meyer, \& Liebert}]{Mamajek:2002cl}
Mamajek, E.~E., Meyer, M.~R., \& Liebert, J. 2002, AJ, 124, 1670

\bibitem[{Marley {et~al.}(2007)Marley, Fortney, Hubickyj, Bodenheimer, \&
  Lissauer}]{Marley:2007bf}
Marley, M.~S., Fortney, J.~J., Hubickyj, O., Bodenheimer, P., \& Lissauer,
  J.~J. 2007, ApJ, 655, 541

\bibitem[{Marley {et~al.}(2012)Marley, Saumon, Cushing, Ackerman, Fortney, \&
  Freedman}]{Marley:2012fo}
Marley, M.~S., Saumon, D., Cushing, M., {et~al.} 2012, ApJ, 754, 135

\bibitem[{Marley {et~al.}(2002)Marley, Seager, Saumon, Lodders, Ackerman,
  Freedman, \& Fan}]{Marley:2002il}
Marley, M.~S., Seager, S., Saumon, D., {et~al.} 2002, ApJ, 568, 335

\bibitem[{Marois {et~al.}(2000)Marois, Doyon, Racine, \&
  Nadeau}]{Marois:2000jt}
Marois, C., Doyon, R., Racine, R., \& Nadeau, D. 2000, PASP, 112, 91

\bibitem[{Marois {et~al.}(2006)Marois, Lafreni{\`e}re, Doyon, Macintosh, \&
  Nadeau}]{Marois:2006df}
Marois, C., Lafreni{\`e}re, D., Doyon, R., Macintosh, B., \& Nadeau, D. 2006,
  ApJ, 641, 556

\bibitem[{Marois {et~al.}(2008)Marois, Macintosh, Barman, Zuckerman, Song,
  Patience, Lafreni{\`e}re, \& Doyon}]{Marois:2008ei}
Marois, C., Macintosh, B., Barman, T., {et~al.} 2008, Science, 322, 1348

\bibitem[{Marois {et~al.}(2010)Marois, Zuckerman, Konopacky, Macintosh, \&
  Barman}]{Marois:2010gp}
Marois, C., Zuckerman, B., Konopacky, Q.~M., Macintosh, B., \& Barman, T. 2010,
  Nature, 468, 1080

\bibitem[{Mathis {et~al.}(1977)Mathis, Rumpl, \& Nordsieck}]{Mathis:1977hp}
Mathis, J.~S., Rumpl, W., \& Nordsieck, K.~H. 1977, ApJ, 217, 425

\bibitem[{McDonald {et~al.}(2012)McDonald, Zijlstra, \&
  Boyer}]{McDonald:2012cg}
McDonald, I., Zijlstra, A.~A., \& Boyer, M.~L. 2012, MNRAS, 427, 343

\bibitem[{McLean {et~al.}(2003)McLean, McGovern, Burgasser, Kirkpatrick, Prato,
  \& Kim}]{McLean:2003hx}
McLean, I.~S., McGovern, M.~R., Burgasser, A.~J., {et~al.} 2003, ApJ, 596, 561

\bibitem[{Meshkat {et~al.}(2013)Meshkat, Bailey, Rameau, Bonnefoy, Boccaletti,
  Mamajek, Kenworthy, Chauvin, Lagrange, Su, \& Currie}]{Meshkat:2013fz}
Meshkat, T., Bailey, V., Rameau, J., {et~al.} 2013, ApJL, 775, L40

\bibitem[{Metchev {et~al.}(2009)Metchev, Marois, \& Zuckerman}]{Metchev:2009jl}
Metchev, S., Marois, C., \& Zuckerman, B. 2009, ApJL, 705, L204

\bibitem[{Mohanty {et~al.}(2007)Mohanty, Jayawardhana, Hu{\'e}lamo, \&
  Mamajek}]{Mohanty:2007er}
Mohanty, S., Jayawardhana, R., Hu{\'e}lamo, N., \& Mamajek, E. 2007, ApJ, 657,
  1064

\bibitem[{Mo{\'o}r {et~al.}(2013)Mo{\'o}r, {\'A}brah{\'a}m, K{\'o}sp{\'a}l,
  Szab{\'o}, Apai, Balog, Csengeri, Grady, Henning, Juh{\'a}sz, Kiss, Pascucci,
  Szul{\'a}gyi, \& Vavrek}]{Moor:2013bg}
Mo{\'o}r, A., {\'A}brah{\'a}m, P., K{\'o}sp{\'a}l, {\'A}., {et~al.} 2013, ApJL,
  775, L51

\bibitem[{Mordasini(2013)}]{Mordasini:2013cr}
Mordasini, C. 2013, A{\&}A, 558, A113

\bibitem[{Mordasini {et~al.}(2012)Mordasini, Alibert, Georgy, Dittkrist, Klahr,
  \& Henning}]{Mordasini:2012gp}
Mordasini, C., Alibert, Y., Georgy, C., {et~al.} 2012, A{\&}A, 547, 112

\bibitem[{Naud {et~al.}(2014)Naud, Artigau, Malo, Albert, Doyon,
  Lafreni{\`e}re, Gagn{\'e}, Saumon, Morley, Allard, Homeier, Beichman, Gelino,
  \& Boucher}]{Naud:2014jx}
Naud, M.-E., Artigau, {\'E}., Malo, L., {et~al.} 2014, ApJ, 787, 5

\bibitem[{Nelder \& Mead(1965)}]{Nelder:1965in}
Nelder, J.~A., \& Mead, R. 1965, The Computer Journal, 7, 308

\bibitem[{Noll {et~al.}(2012)Noll, Kausch, Barden, Jones, Szyszka, Kimeswenger,
  \& Vinther}]{Noll:2012ch}
Noll, S., Kausch, W., Barden, M., {et~al.} 2012, A{\&}A, 543, 92

\bibitem[{{\"O}berg {et~al.}(2011){\"O}berg, Murray-Clay, \&
  Bergin}]{Oberg:2011je}
{\"O}berg, K.~I., Murray-Clay, R., \& Bergin, E.~A. 2011, ApJ, 743, L16

\bibitem[{Partridge \& Schwenke(1997)}]{Partridge:1997kh}
Partridge, H., \& Schwenke, D.~W. 1997, J. Chem. Phys., 106, 4618

\bibitem[{Patience {et~al.}(2010)Patience, King, De~Rosa, \&
  Marois}]{Patience:2010hf}
Patience, J., King, R.~R., De~Rosa, R.~J., \& Marois, C. 2010, 517, A76

\bibitem[{Patience {et~al.}(2012)Patience, King, De~Rosa, Vigan, Witte, Rice,
  Helling, \& Hauschildt}]{Patience:2012cx}
Patience, J., King, R.~R., De~Rosa, R.~J., {et~al.} 2012, 540, A85

\bibitem[{Pecaut {et~al.}(2012)Pecaut, Mamajek, \& Bubar}]{Pecaut:2012gp}
Pecaut, M.~J., Mamajek, E.~E., \& Bubar, E.~J. 2012, ApJ, 746, 154

\bibitem[{Perrin {et~al.}(2014)Perrin, Maire, Ingraham, Savransky,
  Millar-Blanchaer, Wolff, Ruffio, Wang, Draper, Sadakuni, Marois, Rajan,
  Fitzgerald, Macintosh, Graham, Doyon, Larkin, Chilcote, Goodsell, Palmer,
  Labrie, Beaulieu, De~Rosa, Greenbaum, Hartung, Hibon, Konopacky,
  Lafreni{\`e}re, Lavigne, Marchis, Patience, Pueyo, Rantakyr{\"o}, Soummer,
  Sivaramakrishnan, Thomas, Ward-Duong, \& Wiktorowicz}]{Perrin:2014jh}
Perrin, M.~D., Maire, J., Ingraham, P., {et~al.} 2014, Proc. SPIE, 9147, 91473J

\bibitem[{Rameau {et~al.}(2013{\natexlab{a}})Rameau, Chauvin, Lagrange,
  Boccaletti, Quanz, Bonnefoy, Girard, Delorme, Desidera, Klahr, Mordasini,
  Dumas, \& Bonavita}]{Rameau:2013dr}
Rameau, J., Chauvin, G., Lagrange, A.-M., {et~al.} 2013{\natexlab{a}}, ApJL, 772, L15

\bibitem[{Rameau {et~al.}(2013{\natexlab{b}})Rameau, Chauvin, Lagrange,
  Meshkat, Boccaletti, Quanz, Currie, Mawet, Girard, Bonnefoy, \&
  Kenworthy}]{Rameau:2013ds}
---. 2013{\natexlab{b}}, ApJL,
  779, L26

\bibitem[{Rayner {et~al.}(2009)Rayner, Cushing, \& Vacca}]{Rayner:2009ki}
Rayner, J.~T., Cushing, M.~C., \& Vacca, W.~D. 2009, ApJS, 185, 289

\bibitem[{Rizzuto {et~al.}(2011)Rizzuto, Ireland, \&
  Robertson}]{Rizzuto:2011gs}
Rizzuto, A.~C., Ireland, M.~J., \& Robertson, J.~G. 2011, MNRAS, 416, 3108

\bibitem[{Rizzuto {et~al.}(2012)Rizzuto, Ireland, \& Zucker}]{Rizzuto:2012hx}
Rizzuto, A.~C., Ireland, M.~J., \& Zucker, D.~B. 2012, MNRAS, 421, L97

\bibitem[{Rufener(1988)}]{Rufener:1988wq}
Rufener, F. 1988, Sauverny: Observatoire de Geneve, 1988

\bibitem[{Saumon {et~al.}(2000)Saumon, Geballe, Leggett, Marley, Freedman,
  Lodders, Fegley, \& Sengupta}]{Saumon:2000bb}
Saumon, D., Geballe, T.~R., Leggett, S.~K., {et~al.} 2000, ApJ, 541, 374

\bibitem[{Sing {et~al.}(2011)Sing, Pont, Aigrain, Charbonneau, D{\'e}sert,
  Gibson, Gilliland, Hayek, Henry, Knutson, Lecavelier Des~Etangs, Mazeh, \&
  Shporer}]{Sing:2011dn}
Sing, D.~K., Pont, F., Aigrain, S., {et~al.} 2011, MNRAS, 416, 1443

\bibitem[{Skemer {et~al.}(2012)Skemer, Hinz, Esposito, Burrows, Leisenring,
  Skrutskie, Desidera, Mesa, Arcidiacono, Mannucci, Rodigas, Close, McCarthy,
  Kulesa, Agapito, Apai, Argomedo, Bailey, Boutsia, Briguglio, Brusa, Busoni,
  Claudi, Eisner, Fini, Follette, Garnavich, Gratton, Guerra, Hill, Hoffmann,
  Jones, Krejny, Males, Masciadri, Meyer, Miller, Morzinski, Nelson, Pinna,
  Puglisi, Quanz, Quiros-Pacheco, Riccardi, Stefanini, Vaitheeswaran, Wilson,
  \& Xompero}]{Skemer:2012gr}
Skemer, A.~J., Hinz, P.~M., Esposito, S., {et~al.} 2012, ApJ, 753, 14

\bibitem[{Skemer {et~al.}(2014)Skemer, Marley, Hinz, Morzinski, Skrutskie,
  Leisenring, Close, Saumon, Bailey, Briguglio, Defrere, Esposito, Follette,
  Hill, Males, Puglisi, Rodigas, \& Xompero}]{Skemer:2014hy}
Skemer, A.~J., Marley, M.~S., Hinz, P.~M., {et~al.} 2014, ApJ, 792, 17

\bibitem[{Skemer {et~al.}(2015)Skemer, Hinz, Montoya, Skrutskie, Leisenring,
  Durney, Woodward, Wilson, Nelson, Bailey, Defrere, \& Stone}]{Skemer:2015uf}
Skemer, A.~J., Hinz, P., Montoya, M., {et~al.} 2015, eprint arXiv:1508.06290,
  1508.06290

\bibitem[{Skrutskie {et~al.}(2006)Skrutskie, Cutri, Stiening, Weinberg,
  Schneider, Carpenter, Beichman, Capps, Chester, Elias, Huchra, Liebert,
  Lonsdale, Monet, Price, \& Seitzer}]{Skrutskie:2006hla}
Skrutskie, M.~F., Cutri, R.~M., Stiening, R., {et~al.} 2006, AJ, 131, 1163

\bibitem[{Song {et~al.}(2012)Song, Zuckerman, \& Bessell}]{Song:2012gc}
Song, I., Zuckerman, B., \& Bessell, M.~S. 2012, AJ, 144, 8

\bibitem[{Soummer {et~al.}(2012)Soummer, Pueyo, \& Larkin}]{Soummer:2012ig}
Soummer, R., Pueyo, L., \& Larkin, J. 2012, ApJL, 755, L28

\bibitem[{Stephens {et~al.}(2001)Stephens, Marley, Noll, \&
  Chanover}]{Stephens:2001iz}
Stephens, D.~C., Marley, M.~S., Noll, K.~S., \& Chanover, N. 2001, ApJL, 556,
  L97

\bibitem[{Stephens {et~al.}(2009)Stephens, Leggett, Cushing, Marley, Saumon,
  Geballe, Golimowski, Fan, \& Noll}]{Stephens:2009cc}
Stephens, D.~C., Leggett, S.~K., Cushing, M.~C., {et~al.} 2009, ApJ, 702, 154

\bibitem[{Su {et~al.}(2015)Su, Morrison, Malhotra, Smith, Balog, \&
  Rieke}]{Su:2015ju}
Su, K. Y.~L., Morrison, S., Malhotra, R., {et~al.} 2015, ApJ, 799, 146

\bibitem[{Tokunaga {et~al.}(2002)Tokunaga, Simons, \& Vacca}]{Tokunaga:2002ex}
Tokunaga, A.~T., Simons, D.~A., \& Vacca, W.~D. 2002, PASP, 114, 180

\bibitem[{Tokunaga \& Vacca(2005)}]{Tokunaga:2005ch}
Tokunaga, A.~T., \& Vacca, W.~D. 2005, PASP, 117, 421

\bibitem[{van Leeuwen(2007)}]{vanLeeuwen:2007dc}
van Leeuwen, F. 2007, A{\&}A, 474, 653

\bibitem[{Wang {et~al.}(2015)Wang, Ruffio, De~Rosa, Aguilar, Wolff, \&
  Pueyo}]{Wang:2015th}
Wang, J.~J., Ruffio, J.-B., De~Rosa, R.~J., {et~al.} 2015, Astrophysics Source
  Code Library, -1, 06001

\bibitem[{Wang {et~al.}(2014)Wang, Rajan, Graham, Savransky, Ingraham,
  Ward-Duong, Patience, De~Rosa, Bulger, Sivaramakrishnan, Perrin, Thomas,
  Sadakuni, Greenbaum, Pueyo, Marois, Oppenheimer, Kalas, Cardwell, Goodsell,
  Hibon, \& Rantakyr{\"o}}]{Wang:2014ki}
Wang, J.~J., Rajan, A., Graham, J.~R., {et~al.} 2014, Proc. SPIE, 9147, 55

\bibitem[{Witte {et~al.}(2011)Witte, Helling, Barman, Heidrich, \&
  Hauschildt}]{Witte:2011kn}
Witte, S., Helling, C., Barman, T., Heidrich, N., \& Hauschildt, P.~H. 2011,
  A{\&}A, 529, A44

\bibitem[{Woitke \& Helling(2003)}]{Woitke:2003cs}
Woitke, P., \& Helling, C. 2003, A{\&}A, 399, 297

\bibitem[{Woitke \& Helling(2004)}]{Woitke:2004ie}
---. 2004, A{\&}A, 414, 335

\bibitem[{Wolff {et~al.}(2014)Wolff, Perrin, Maire, Ingraham, Rantakyr{\"o}, \&
  Hibon}]{Wolff:2014cn}
Wolff, S.~G., Perrin, M.~D., Maire, J., {et~al.} 2014, Proc. SPIE, 9147, 91477H

\bibitem[{Wright {et~al.}(2003)Wright, Egan, Kraemer, \& Price}]{Wright:2003gs}
Wright, C.~O., Egan, M.~P., Kraemer, K.~E., \& Price, S.~D. 2003, AJ, 125, 359

\bibitem[{Wright {et~al.}(2010)Wright, Eisenhardt, Mainzer, Ressler, Cutri,
  Jarrett, Kirkpatrick, Padgett, McMillan, Skrutskie, Stanford, Cohen, Walker,
  Mather, Leisawitz, Gautier, McLean, Benford, Lonsdale, Blain, Mendez, Irace,
  Duval, Liu, Royer, Heinrichsen, Howard, Shannon, Kendall, Walsh, Larsen,
  Cardon, Schick, Schwalm, Abid, Fabinsky, Naes, \& Tsai}]{Wright:2010in}
Wright, E.~L., Eisenhardt, P. R.~M., Mainzer, A.~K., {et~al.} 2010, AJ, 140,
  1868

\bibitem[{Zahnle \& Marley(2014)}]{Zahnle:2014hl}
Zahnle, K.~J., \& Marley, M.~S. 2014, ApJ, 797, 41

\bibitem[{Zhou {et~al.}(2014)Zhou, Herczeg, Kraus, Metchev, \&
  Cruz}]{Zhou:2014ct}
Zhou, Y., Herczeg, G.~J., Kraus, A.~L., Metchev, S., \& Cruz, K.~L. 2014, ApJ,
  783, L17

\bibitem[{Zurlo {et~al.}(2013)Zurlo, Vigan, Hagelberg, Desidera, Chauvin,
  Almenara, Biazzo, Bonnefoy, Carson, Covino, Delorme, D'Orazi, Gratton, Mesa,
  Messina, Moutou, Segransan, Turatto, Udry, \& Wildi}]{Zurlo:2013kb}
Zurlo, A., Vigan, A., Hagelberg, J., {et~al.} 2013, A{\&}A, 554, 21

\end{thebibliography}
\end{document}